%% file: template.tex
\PassOptionsToPackage{svgnames}{xcolor}
\documentclass[journal]{vgtc} % review (journal style)
\onlineid{1773}

\vgtccategory{Research}

\vgtcpapertype{application}


\title{ClimateSOM: A Visual Analysis Workflow\\ for Climate Ensemble Datasets}

\author{
\authororcid{Yuya Kawakami}{0000-0002-8621-301X}, 
\authororcid{Daniel Cayan}{0000-0002-2719-6811}, 
\authororcid{Dongyu Liu}{0000-0002-8915-2785}, and
\authororcid{Kwan-Liu Ma}{0000-0001-8086-0366}
}
\authorfooter{
  \item
  	Yuya Kawakami, Dongyu Liu, and Kwan-Liu Ma are with University of California, Davis.
  	E-mails: \{ykawakami, dyuliu, klma\}@ucdavis.edu
  \item
  	Danial Cayan is with Scripps Institution of Oceanography, University of California, San Diego.
  	E-mail: dcayan@ucsd.edu
}

\input{tex/0_abstract}

\graphicspath{{figs/}{figures/}{pictures/}{images/}{./}} % where to search for the images

\usepackage{mathptmx}                  % use matching math font
\usepackage[color=purple!15]{todonotes}
\usepackage{algorithm,algpseudocode}
\usepackage{array}
\usepackage{amssymb}
\usepackage{xspace}
\usepackage{listings}
\usepackage{amsmath}
\usepackage[breakable]{tcolorbox}
\usepackage[pagebackref,bookmarks]{hyperref}
\usepackage[nameinlink,capitalise]{cleveref}
\usepackage{float}
\usepackage[normalem]{ulem}
\newcommand{\ClimateSOM}{{\fontfamily{zi4}\selectfont ClimateSOM}\xspace}

\newcommand{\highlight}[1]{\colorbox{gray!120}{\textbf{\textcolor{white}{\footnotesize #1}}}}

\definecolor{navy}{RGB}{0, 0, 100}
\definecolor{lightblue}{RGB}{199, 233, 255}
\definecolor{lightgreen}{RGB}{185, 255, 185}
\definecolor{circuitred}{RGB}{239, 37, 21}
\newcommand{\subimage}[1]{\tcbox[on line, size=minimal, boxrule=0pt, colback=navy!100,left=3pt,right=3pt,top=1pt,bottom=1pt, arc=4pt, frame empty]{\textcolor{white}{\footnotesize #1}}}

\newcommand{\req}[1]{\tcbox[on line, size=tight, boxrule=0pt, colback=lightblue!100,left=3pt,right=3pt,top=2pt,bottom=2pt, arc=5pt, frame empty ]{\textcolor{black}{\textbf{#1}}}}
\newcommand{\dist}[1]{\tcbox[on line, size=tight, boxrule=0pt, colback=circuitred!100,left=2pt,right=2pt,top=2pt,bottom=2pt, arc=3pt, frame empty ]{\textcolor{white}{#1}}}
\newcommand{\step}[1]{\tcbox[on line, size=tight, boxrule=0pt, colback=lightgreen!100,left=3pt,right=3pt,top=1pt,bottom=1pt, arc=0pt, frame empty ]{\textcolor{black}{\textbf{#1}}}}

\makeatletter
\renewcommand{\paragraph}{%
  \@startsection{paragraph}{4}{0pt}{0pt}{-1em}{\normalfont\normalsize\bfseries}
}
\makeatother
\newif\ifdiffvisible
\diffvisiblefalse
\definecolor{diffdel}{RGB}{255,0,0}
\definecolor{diffadd}{RGB}{0,128,0}
\newcommand\revdel[1]{\ifdiffvisible{\color{diffdel}\sout{#1}}\fi}
\newcommand\revadd[1]{\ifdiffvisible{\color{diffadd}#1}\else#1\fi}

\begin{document}

%%%%%%%%%%%%%%%%%%%%%%%%%%%%%%%%%%%%%%%%%%%%%%%%%%%%%%%%%%%%%%%%
%%%%%%%%%%%%%%%%%%%%%% START OF THE PAPER %%%%%%%%%%%%%%%%%%%%%%
%%%%%%%%%%%%%%%%%%%%%%%%%%%%%%%%%%%%%%%%%%%%%%%%%%%%%%%%%%%%%%%%

\firstsection{Introduction}
\label{sec:intro}
\maketitle

\input{tex/1_intro}

\section{Related Works}\label{sec:relworks}
\input{tex/2_relworks}

\section{Problem Characterization}\label{sec:problem}
\input{tex/3_problem}
\section{ClimateSOM}\label{sec:workflow}
\input{tex/4_workflow}
\section{The Visual Interface}

\input{tex/5_interface}

\section{Evaluation}
\input{tex/6_usecases}
\section{Expert Feedback}
\input{tex/7_feedback}

% \section{Discussion}
% \input{tex/8_discussion}
% \vspace{-5pt}
\section{Conclusion}
\input{tex/9_conclusion}
%% if specified like this the section will be omitted in review mode
\acknowledgments{%
	This research is supported in part by the National Science Foundation via grant No. IIS-2427770}
    
    % and by the UC Climate Action Initiative grant.
    % }

% \bibliographystyle{abbrv-doi-hyperref}
\bibliographystyle{abbrv-doi-hyperref-narrow}

\bibliography{template}
\clearpage
\appendix
\input{tex/appendix}

\end{document}

%% file: tex/0_abstract.tex
\abstract{
Ensemble datasets are ever more prevalent in various scientific domains.
In climate science, ensemble datasets are used to capture variability in projections under plausible future conditions including greenhouse and aerosol emissions.
Each ensemble model run produces projections that are fundamentally similar yet meaningfully distinct.
Understanding this variability among ensemble model runs and analyzing its magnitude and patterns is a vital task for climate scientists.
In this paper, we present \ClimateSOM, a visual analysis workflow that leverages a self-organizing map (SOM) and Large Language Models (LLMs) to support interactive exploration and interpretation of climate ensemble datasets.
The workflow abstracts climate ensemble model runs---spatiotemporal time series---into a distribution over a 2D space that captures the variability among the ensemble model runs using a SOM.
LLMs are integrated to assist in sensemaking of this SOM-defined 2D space, the basis for the visual analysis tasks.
In all, \ClimateSOM enables users to explore the variability among ensemble model runs, identify patterns, compare and cluster the ensemble model runs.
To demonstrate the utility of \ClimateSOM, we apply the workflow to an ensemble dataset of precipitation projections over California and the Northwestern United States.
Furthermore, we conduct a short evaluation of our LLM integration, and conduct an expert review of the visual workflow and the insights from the case studies with six domain experts to evaluate our approach and its utility.
}

%% file: tex/1_intro.tex
Ensemble datasets are increasingly integral to scientific research as computing resources have increased in power \cite{wang_visualization_2019, kehrer2013visualization}. 
Many disciplines, from computational fluid dynamics \cite{patchett2017deep} to material science\cite{wang_visualization_2019}, rely on ensemble datasets to capture and predict variability by running their models under different conditions and parameters.
The result is a collection of model runs---an \textit{ensemble} dataset---that supports exploration of possible outcomes, sensitivity analysis, and uncertainty quantification. However, working with ensemble datasets remains difficult due to their sheer size and complexity~\cite{wang_visualization_2019}.
% The output of such a process, a set of many model runs known as an \textit{ensemble} dataset, helps researchers explore the range of possible outcomes, understand the sensitivity of the outputs with respect to the model variants or parameters, and quantify uncertainty in predictions. Despite its ubiquity, a major challenge with working with ensemble datasets is often their sheer size and complexity \cite{wang_visualization_2019}.

% Since a single model run often requires a specialized analysis, this difficulty only grows when dealing with an entire ensemble.
% Since analyzing a single output set from a particular  model run often requires individual analysis, dealing with an entire ensemble only amplifies this difficulty.
% They often exhibit substantial variability both within and across model runs, making them difficult for domain scientists to fully interpret them.
% ensemble datasets often highly vary within and across model runs, 

Climate science in particular relies on ensemble datasets to examine plausible climate futures under varying conditions, such as greenhouse gas and aerosol emissions or land use changes
~\cite{murphy_quantification_2004, eyring2016overview, maher_max_2019}.
Each model run often represents  a spatiotemporal time series, where each time step is a climate outcome (e.g., precipitation) over some spatial regions.
These model runs may come from different global climate models~\cite{eyring2016overview, maher_max_2019} and use varying input assumptions, producing a broad set of outcomes. Understanding and communicating the resulting variability and patterns remains a major challenge.

Although existing analyses in climate science provide valuable insights---often through statistical summaries like spatial means and variances across model runs---they may overlook key spatial and temporal structures~\cite{kappe_exploring_2019}.
Visual analytics techniques have been developed to support interpretation of ensemble data~\cite{biswas2016visualization, kappe_exploring_2019, evers_interactive_2023}, but many struggle to provide both interpretability and robustness when dealing with highly complex, chaotic model behavior.
%
% The main challenge, particularly with climate ensemble datasets, as previously identified\cite{potter2009visualization}, is that model runs are often both highly chaotic and complex.
%
% In addition to the systematic variations introduced by the model inputs that climate scientists are interested in exploring, they usually also include unknown amounts of random variation.
% In addition to the systematic climate variations driven by the model inputs that are sought after, any analytics and must contend with unknown amounts of random variation.
% they include unknown amounts of random variation in addition to the systematic variations of interest introduced by the GCMs and SSPs.
%
% In face of such variability and complexity, 
%
% As Kappe outlines, in the context of ensemble spatiotemporal simulations, ``direct visualizations \textit{(of climate ensembles)} are prohibitive with respect to space, time, and mental capacity of the user'' \cite{kappe2022visual}. 
%
% Simple exploratory visualizations of the this data often reveal that the innumerable number of variations observed in the dataset is too complex to understand or visualize in its entirety.
In this context, even summarizing a single model run clearly is difficult, and comparing or clustering multiple runs poses a greater challenge.

% As a result, most approaches resort to \textit{Summary statistics} 

To overcome these limitations, we introduce, \ClimateSOM, a new visual analysis workflow that equips climate scientists with an intuitive tool and visual language for exploring and analyzing ensemble climate projection datasets. 
In line with findings by Wang et al. in their comprehensive survey of ensemble data visualization \cite{wang_visualization_2019}, we identify three essential tasks when working with such datasets: \textit{(1) exploring a single set of model run(s)}, \textit{(2) comparing the behavior across two sets of model runs}, and \textit{(3) clustering the behavior of multiple model runs}. 
These tasks are crucial for uncovering insights into climate variability and making meaningful inferences about these projections. 
These tasks along with its interpretability serves as \revdel{a}\revadd{our} guiding principle.\revdel{for \ClimateSOM}

The overarching goal for \ClimateSOM is data abstraction by transforming spatiotemporal time series into a distribution over a user-steerable 2D space.
Central to \ClimateSOM is a self-organizing map (SOM)\cite{kohonen_self-organizing_1990}, a widely recognized analytical tool in climate science, along with steerable dimensionality reduction and annotations.
%
% Expanding on past work on SOM training, we highlight the impact of SOM training parameters and its importance to the success of this visual workflow.
%
Through this transformation, \ClimateSOM enables flexible, interpretable exploration of diverse model behavior, helping researchers detect patterns and relationships that would otherwise be difficult to uncover.

We further integrate LLMs to assist in sensemaking of the SOM-defined 2D space.
\ClimateSOM introduces an LLM-empowered bidirectional query methods that allows users to (1) locate regions of interest in the 2D space based on natural language queries and (2) generate textual summaries of selected regions.
We demonstrate the utility of \ClimateSOM through case studies using LOCA-downscaled CMIP6 climate projections~\cite{pierce2023future} across two regions in the western United States. In addition, we conduct an expert review with six domain scientists to assess the system’s usability and its ability to support scientific insight.
In short, the contributions of this work are the following:
\begin{itemize}[noitemsep, topsep=0pt, leftmargin=8pt ]
    \item A novel workflow for visual analysis of a spatiotemporal ensemble that abstracts the ensemble members as a distribution over a steerable 2D space framed by a self-organizing map.
    \item Integration of LLMs to assist in sensemaking of the SOM-defined 2D space in which \ClimateSOM operates.
    \item \ClimateSOM, a visualization interface that supports this workflow end-to-end, enabling users to explore, compare, and cluster ensemble model runs.
    % \item Case studies on the LOCA-Downscaled CMIP6 climate model projections over two different regions of the Western United States and interviews with six domain experts to evaluate its efficacy.
        % to demonstrate the utility of \ClimateSOM in uncovering insights from climate ensemble datasets.
\end{itemize}

%% file: tex/2_relworks.tex
%In their survey, Wang et al. identifiy five \textit{dimensions} of ensemble data: variable, location, time, member, and ensemble. 
%%
%They also define six \textit{analytic tasks} for ensemble data: overview, compare, cluster, trend, feature, and parameters.
%%
%Each \textit{analytic task}, when applied to different \textit{dimensions}, demands a distinct analysis and visualization strategy\cite{wang_visualization_2019}.
% For any of the \textit{analytic tasks}, the \textit{dimension} considered brings rise to a entirely different analysis and visualization strategy \cite{wang_visualization_2019}.
%
% We adopt this taxonomy to structure our discussion of the visualization strategies for ensemble climate data.

\subsection{Visual Analytics for Climate Ensemble Data}
% \subsubsection*{\note{Proposed Structure}}

% \begin{tightItemize}
%     \item Here are past work that have tried climate ensemble data visualization.
%     \item Given the complexity of this data, visualization and analytics of such data often rely on data abstraction and aggregation to compress the data into a concise form.
%     \item For climate ensemble data, the most straightforward intermediate data abstraction are summary statistics as is populate in climate science for its simplicity and interpretablilty\cite{kehrer_brushing_2010,hollt_ovis_2014}. However, past work have also considered graph-based approach, where a graph is constructed to represent the ensemble \cite{shu2016ensemblegraph, kappe_analysis_2019}, topology-based methods, where isocontours, spaghetti plots or similar methods are employed to represent the ensemble data while retainig the spatial context,  \cite{kappe_topology-based_2019,kappe_topology-based_2022, liu_visualizing_2015, vietinghoff_visual_2021, sanyal_noodles_2010, ma_interactive_2019}, or even as 2D images\cite{ise_varenn_2020}.

% \end{tightItemize}

% Given the abundance of ensemble data sets from large-scale simulations, many visualization techniques to investigate the ensemble data have been proposed in literature \cite{potter2009ensemble}.

The increasing availability of ensemble data from large-scale climate simulations has led to the development of a range of visual analytics methods designed to manage and interpret uncertainty, spatial-temporal patterns, and inter-model variability~\cite{kehrer_brushing_2010, biswas_visualization_2017, ferstl_time-hierarchical_2017, wang_multi-resolution_2017,biswas2016visualization, kappe_exploring_2019,bensema2016modality,hollt2014ovis}. 
%
% Given the complexity of this data, visualization and analytics of such data often rely on data abstraction and aggregation to compress the data into a concise form.
%
% For climate ensemble data, the most common data abstraction is summary statistics as is popular in climate science for its simplicity and interpretablilty \cite{kehrer_brushing_2010,hollt_ovis_2014}. 
%
% Of the many methods, statistics~\cite{kehrer_brushing_2010, hollt_ovis_2014} and matrix decomposition~\cite{chandler_characterizing_2024} have been central tools for some previous works.
%
Past works have considered topology-based methods, where isocontours, spaghetti plots or similar methods are employed to \revdel{preserve spatial context}\revadd{encapsulate prominent spatial features} while encoding ensemble variability~\cite{kappe_topology-based_2019,kappe_topology-based_2022, liu_visualizing_2015, sanyal_noodles_2010, ma_interactive_2019}.
Others have proposed graph-based representations of ensemble relationships~\cite{shu2016ensemblegraph, kappe_analysis_2019} or treated ensemble members as collections of images to facilitate comparative analysis~\cite{ise_varenn_2020}.
%
% In \cite{shu2016ensemblegraph} Shu et al. presented EnsembleGraph, a visual analytics tool for time-varying ensemble data that abstracted spatial regions as nodes to develop a graph-based representation of the ensemble.
%
Kappe et al. introduced a Sankey-diagram-based visualization using \textit{k}-means clustering on the entire climate field to summarize variability in decadal predictions~\cite{kappe_exploring_2019}, while Evers et al. employed spatial correlation measures and multidimensional scaling (MDS) to embed ensemble members for comparative visualization~\cite{evers_uncertaintyaware_2021}.
%
% This work was extended replacing hierarchical clustering with \textit{k}-means clustering \cite{kappe_analysis_2019}.
% to build a more advanced visual analytics tool 
%
% More recently, Evers et al. demonstrated a visualization workflow for climate ensemble via calculating spatial correlation across the spatial domain and then embedding the climate ensemble using multi-dimensional scaling (MDS)~\cite{evers_uncertaintyaware_2021}.
%
% This method allows users to understand regions of spatial domain that exhibit similar behavior across the ensemble members.
%
% This work was further extended for multi-field climate ensembles by calculating spatial correlations on a concatenated time series across all fields~\cite{evers_interactive_2023}.

%In both set of works, clustering of the ensemble data is identified as a key \textit{analytic task}, though they differ the \textit{dimension}; the former clusters the ensemble members, while the latter clusters the spatial domain.
%%
%\ClimateSOM aligns closer to \cite{kappe_analysis_2019, kappe_exploring_2019} where the clustering on the ensemble members is the primary focus.

% Given the complexity of climate ensemble data, the balance between the level of abstraction and the amount of information retained is a key challenge in its visualization, and we propose \ClimateSOM as another approach to address this challenge, this time to abstract the data into a 2D distribution.

However, in the context of visualizing the overall behavior of a single climate ensemble model run or its comparison, past works have mostly relied on summary statistics \revadd{of the climate field interest} or what Kappe refers to the as the \textit{mean field approach}~\cite{kappe2022visual}.
\revdel{As Kappe points out, however, this aggregation can be misleading since climate ensemble data seldom vary around a single mean, instead exhibiting multiple means of varying importance and distributional characteristics.}
\revadd{Such approaches, as well as topology-based methods that often select most prominent spatial features to represent ensemble members for analysis and comparison \cite{ferstl_time-hierarchical_2017}, may crucially miss out of the different distributional nature of the ensemble members, both within one member and across members.}

Departing from this \textit{mean field} aggregation is a key motivation for \ClimateSOM. Rather than reducing the ensemble to a single statistical summary \revadd{or a topological feature}, our approach models it as a 2D spatially organized distribution, preserving the variability with each model run.
\revadd{This allows us to capture fundamental differences within and across ensemble members that may be ignored under the \textit{mean field approach}, while allowing us to succinctly formalize notions of comparison and clustering within the workflow.} 
We use a \revdel{self-organizing map (SOM)}\revadd{SOM} to maintain local topological relationships while revealing fine-grained differences between ensemble members. In contrast to prior systems, which typically support only isolated subtasks, such as single-run exploration or pairwise comparison, \ClimateSOM provides integrated support for exploration, comparison, and clustering. This unified capability addresses a gap in existing climate ensemble visual analytics.

%\subsection{Dimensionality Reduction (DR) for Spatiotemporal Data}
%The manifold hypothesis \cite{fefferman2016testing} that posits that high-dimensional data lies on a low-dimensional manifold has been a driving force behind the development of dimensionality reduction (DR) techniques.
%%
%Notato and Aupetit provide a comprehensive survey of DR or projection methods for visual analytic tasks \cite{nonato_multidimensional_2019}.
%%
%Spatial or spatiotempoiral data is no exception to this and DR methods can be readily to spatial data as well, providing an avenue to visually explore such data.
%%
%Classical Multidimensional Scaling (MDS) is often used in these works \cite{piccolotto_undrground_2024} \note{Cite more ST-MDS works.}
%%
%In \cite{piccolotto_undrground_2024}, Piccolotto et al. used MDS to visualize latent components and their relationships that were extratcted from a Spatial Blind Source Separation (SBSS) analysis.

%%
%As Nonato and Aupetit outline the in any DR or similar projection method, the choice of the dissilarity measure is crucial to the success of the method.
%%
%Although Euclidean distance is the most common choice, its expressive power is limited in cases where the distance is large as in images \cite{agrawal2021minimum}.
%%
%Kappe explored the role of dissimilarity measures in the context of climate data \cite{kappe2022visual}

%Brehmer et al. identified a number of common tasks for visualization with dimensionality reduction, including the intepretation of the synthesized dimension \cite{brehmer_visualizing_2014}.

\subsection{SOMs in Visual Analytics and Climate Science}
Self-organizing maps (SOMs)~\cite{kohonen_self-organizing_1990} are commonly used for their ability to project high-dimensional data onto a structured, low-dimensional grid while preserving local topological relationships. This pseudo-continuous 2D layout enables spatial interpretation and supports visual pattern recognition.
In visual analytics, SOMs have been used to structure time-series and spatial data, as demonstrated by Andrienko et al. in their analysis of crime patterns ~\cite{andrienko_space-time_2010}. 
Beyond this, SOMs have been applied in diverse domains such as astrophysics and cytometry, where the preservation of local structure facilitates cluster identification and improves interpretability~\cite{chen_self-organizing_2013, hemmati_bringing_2019, sacha_somflow_2018, chushig-muzo_data-driven_2020, van_gassen_flowsom_2015,quintelier_analyzing_2021, tu_phrasemap_2024, eirich_irvine_2022}.

In climate science, SOMs have often been used for pattern extraction, where extracted nodes represent particular spatial patterns that may be linked to known physical phenomena \cite{liu_performance_2006,reusch_north_2007,mihanovic_surface_2011}.
% The ``\textit{self-organizing}'' property of SOMs, which enables dimensionality reduction, is often credited as a key feature that makes analysis via SOMs fruitful in Climate Sciences, where extracted nodes and patterns can often be associated with known physical phenomena \cite{liu_performance_2006}.
%
\revdel{For instance, Reusch et al. used SOMs to identify the North Atlantic Oscillation (NAO) in climate data \cite{reusch_north_2007}, while Mihanovi\'{c} et al. used SOMs to identify surface circulation patterns in the Northern Adriatic Sea \cite{mihanovic_surface_2011}.}
%
% Nourani et al. used SOMs to preprocess precipitation satellite data for further analysis \cite{nourani_using_2013} and Loikith et al. used SOMs to identify large-scale meteorological patterns (LSMPs) over the Northwestern United States \cite{loikith_characterizing_2017}.
%
For more applications of SOMs in climate science or related areas, we refer the reader to \cite{liu_review_2011, sheridan_self-organizing_2011, hewitson_self-organizing_2002, gibson_use_2017}.
Regardless of the exact application, the trained SOM and its associated 2D node space must be interpretable by the user to be useful as this serves as the backbone for its use.
Past works have mostly considered Sammon Mapping or Multidimensional Scaling (MDS) to understand the SOM-defined 2D space; however, these methods do not provide user control over the mapping process.

In \ClimateSOM, we extend SOM-based approaches by integrating interactive visual analytics techniques. In particular, we augment the SOM node space with interpretability tools including user annotations, interactive dimensionality reduction, and LLM-empowered bidirectional queries. This allows domain scientists to both guide and interrogate the structure learnt by the SOM, moving beyond static representations and improving the transparency and utility of the latent 2D embedding.

\subsection{LLMs for VIS}
Natural Language Interfaces (NLI)~\cite{shen_towards_2023} or LLMs have been increasingly applied to assist in visualization workflows~\cite{hutchinson_llm_2024, zhang_natural_2024, zhao_leva_2025}.
Such systems show promise in helping users by generating interpretable annotations in visualizations or by providing a conversational interface to the visualization system~\cite{wu_chartinsights_2024,shen_towards_2023} which can aid in their sensemaking process~\cite{bernard_comparing_2018,rahman_qualitative_2025}.
They have also been shown to alleviate the ``cold-start problem'' in information-seeking, helping users establish context and find a starting point when navigating new information~\cite{liu_selenite_2024}.
LLMs have also been successfully applied for Text-to-SQL tasks ~\cite{zhang_natural_2024, li_dawn_2024} and for summarizing insights from and improve usability in visual analytics interfaces~\cite{zhao_leva_2025}.
In fact, LLMs have been used in geospatial visual analytics systems as in \cite{li_save_2024} for gathering and summarizing insights in a geospatial context.
% or for generating SQL queries from natural language questions
%

In \ClimateSOM, we integrate LLMs more tightly into the system design by enabling bidirectional interaction between users and the SOM-defined latent space. This includes support for text-based queries into the embedding and automated summarization of regions in the latent space. This approach not only improves usability but also enhances the interpretive bandwidth of the SOM, particularly in settings where exploratory context is limited.

% With \ClimateSOM, we enable users to provide \textit{annotations} in the SOM-defined 2D space as a means to express their information needs, leveraging LLMs in the process.
%
% Meanwhile, the latent space of dimensionality reduction techniques have been freqeuntly explored in the past \cite{sacha_visual_2017}, many visual analytics works seeking to improve its interpretablity \cite{montambault_dimbridge_2025, eckelt_visual_2023}.
%
% Despite the documented issues like hallucinations that LLMs face, in light of these works, we propose the use of LLMs to help users navigate the latent space of the SOM in the \ClimateSOM workflow and reduce the "cold start" problem in exploring this latent space.
%Visualizing complex data often involves  
%dimensionality reduction to a lower-dimensional space, usually in 2D, to facilitate its understanding.
%%
%The premise is that patterns in the data are more easily discernible after processing; however, the problem of understanding the patterns still remains albeit in a lower-dimensional space.
%%
%Recent works have explored the use of LLMs to assist in sensemaking in these latent spaces \cite{raval_explain-and-test_2023}.

%% file: tex/3_problem.tex
We focus on ensemble datasets such as those produced by the Coupled Model Intercomparison Project (CMIP6)~\cite{eyring2016overview}, which contain climate projections generated under a variety of input assumptions. Each simulation produces a spatiotemporal time series, referred to as a model run. The following terminology is used throughout:
\begin{itemize}[noitemsep, topsep=0pt, leftmargin=8pt]
    \item \textbf{GCM} (Global Climate Model): Mathematical models that simulate the Earth's climate system, and are used to project future climate scenarios (e.g., MIROC6, ACCESS-CM2, etc.).
    \item \textbf{SSP} (Shared Socioeconomic Pathways)~\cite{riahi2017shared}: Pathways that examine how global society, demographics, and economics might change over the next century. 
    In addition to the historical period, we consider three SSPs in this work: SSP245, SSP370, and SSP585, a low, medium high, and high greenhouse gas emissions future respectively. 
    \item \textbf{Model run}: A unique climate projection and spatiotemporal time series, representing a member of the climate ensemble generated by a specific combination (GCM, SSP). The full ensemble consists of multiple such runs.
\end{itemize}

\subsection{Design Goals}\label{ssec:design_requirements}
In the project, in consultation with a climate scientist (paper co-author) specializing in regional climate variability and changes, we identified three key questions that drive exploratory analysis of ensemble climate datasets, which informs our system's design goals.
% Given a climate ensemble dataset, we list the motivating questions and the design requirements for \ClimateSOM.
\begin{itemize}[left=0pt, itemsep=0pt, labelsep=3pt, align=left,  itemindent=0pt]
    \item[\req{R1}] {\normalfont \sffamily``Q: What is the general behavior of the set A of model runs?''}  \\ A climate scientist wishes to understand the general behavior of a single model run or a set of model runs and its general tendencies as it relates to the climate projection it generates. For example, do the model run(s) consistently produce projection of a certain quality or does it vary over time? \\ $\rightarrow$ \textbf{Requirement: }\textit{Provide a visual language \revdel{and description} to explore the behavior of a single set of model runs over the entire temporal domain.}
    \item[\req{R2}] {\normalfont \sffamily``Q: How does set A of model runs compare to set B of model runs?''} \\ A climate scientist wishes to compare the behavior of two sets of model runs A and B to understand the differences in their projections. For example, does set A of model runs produce similar projections compared to set B of model runs? What overlapping qualities do they share and what climate projection qualities are unique to \revdel{run A or run B}\revadd{each}?\\ $\rightarrow$ 
    \textbf{Requirement: }\textit{Provide a visual language \revdel{description and support} to compare the behavior of two sets of model runs over the entire temporal domain.}
    \item[\req{R3}]{\normalfont \sffamily``Q: Which model runs or GCMs exhibit \textit{similar} behavior?''} \\ A climate scientist wishes to understand whether there are meaningful clusterings of ensemble runs and clustering of GCMs based on the SSP forcings. Do all ensemble runs produce similar results? Is the effect of SSP forcing similar on all GCMs?\\ $\rightarrow$ 
    \textbf{Requirement: }\textit{Support clustering of ensemble model runs based and clustering of GCM based on their SSP forcings over the entire temporal domain.} 
\end{itemize}
% The goal of \ClimateSOM is to provide answers to the above question while ensuring that they are readily decipherable.
% The goal of \ClimateSOM is to address the above questions with an interpretable visualization tool.
% \note{How to properly incorporate interpretablitity as a design requirement?}
% As outlined in~\cref{sec:intro} the main task

\subsection{Abstracting the Model Runs Into a Distribution}\label{ssec:abstraction}
\begin{figure}[H]
   \includegraphics[width=\linewidth]{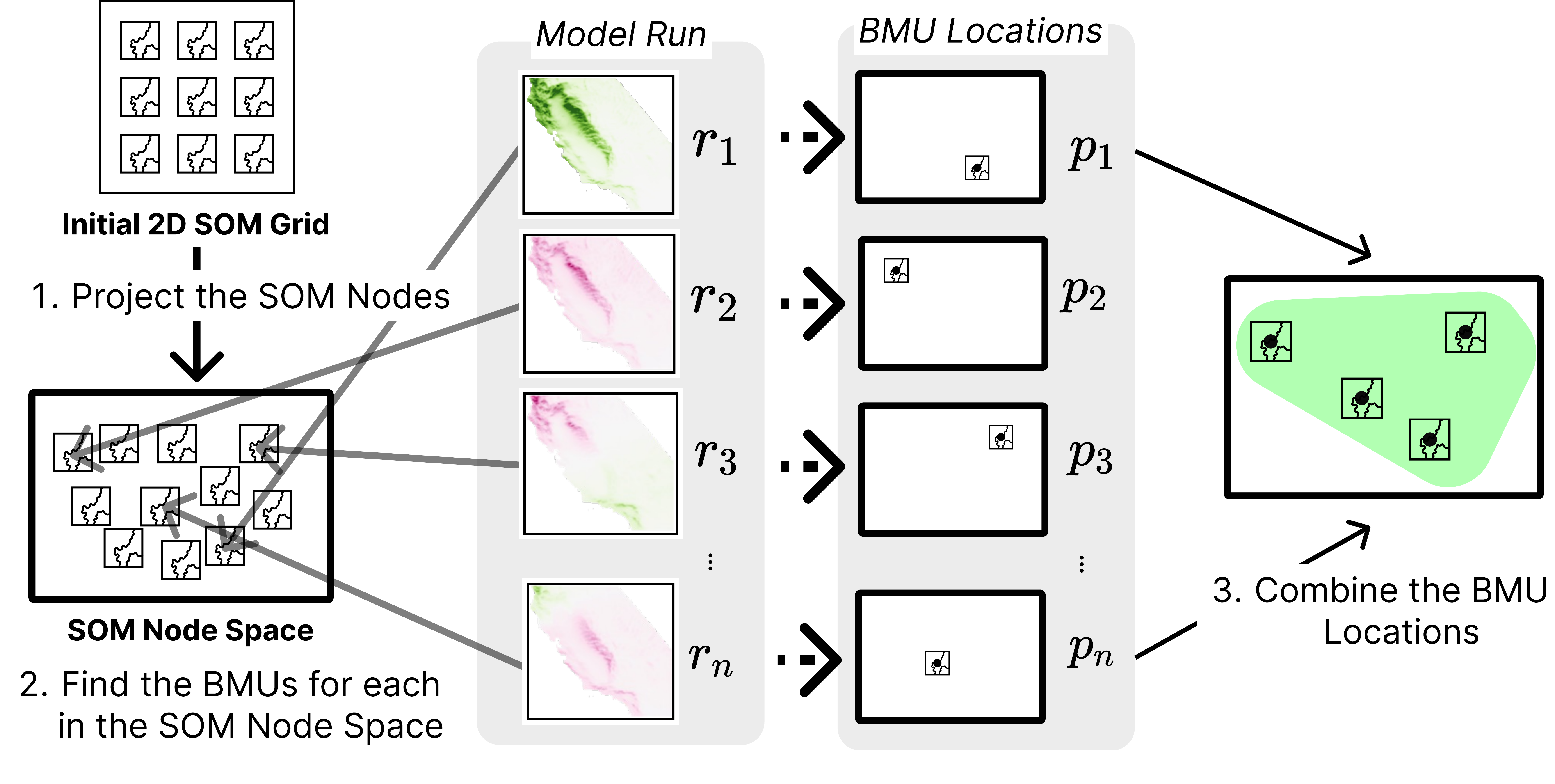}
    \caption{The distributional abstraction for model runs in \ClimateSOM. Each model runs are abstracted to a distribution defined by a projected location of SOM Nodes, the \textit{SOM Node Space}.}
    \vspace{-0.1in}
    \label{fig:SOM}
\end{figure}

Central to \ClimateSOM is a self-organizing map (SOM) trained to capture the spatial variability of a climate ensemble dataset, while retaining its interpretability.
An SOM is an unsupervised neural network that projects high-dimensional data into a 2D grid while preserving the topological structure, making it \revdel{a useful tool}\revadd{useful} for dimensionality reduction (DR) \revdel{of its input}.
Training a SOM over a climate ensemble dataset yields a 2D SOM grid, with each SOM node representing a spatial pattern over the studied region.
\ClimateSOM leverages this capability to represent each model run within the ensemble as a distribution.
Specifically, for each time step we first locate its best matching unit (BMU) in the SOM nodes, and allow the 2D position of that SOM node to represent that time step.
Aggregating these positions across all time steps produces a 2D distribution defined by the set of BMU's node locations.
\cref{fig:SOM} presents the construction of the distributional representation.

Noteworthy in this construction, however, is the projection step of the initial 2D SOM grid. 
While the SOM nodes maintain local smoothness, they can have distortions globally~\cite{sheridan_self-organizing_2011}; thus, defining the SOM node locations solely by their 2D grid location is not desirable.
To address this, we add a projection step of the SOM nodes to better preserve their relative structure and enhance its interpretability.
% 
%We thus aim to correct for these distortions by projecting the SOM nodes into a more interpretable orientation to better preserve the relative structure of the SOM nodes.
%
% Finally, we use the projected SOM node locations to construct the distributional representation as shown in~\cref{fig:SOM}.
\noindent
Formally, for an ensemble run with $n$ time steps, each of which a climate projection over some spatial domain, $R=[r_1, r_2, \ldots, r_n]$, we abstract this as a set of points in 2D space:
% \vspace{-4pt}
\begin{align*}
S=\{p_1,p_2,\ldots,p_n\}, \,\,\, p_i=\text{proj}(BMU(r_i))
\end{align*}
% \vspace{-4pt}
where $BMU(r_i)$ is the best matching SOM node for $r_i$, and proj$(\centerdot)$ maps the node to a 2D layout.
This distributional representation of the model runs serves as the backbone of \ClimateSOM.
%
% Therefore, enabling the interpretation of the projected space of the SOM nodes is principal to our workflow which we

% as a set of points $S=\{p_1, p_2, \ldots, p_n\}$ in an interpretable 2D space, where each $p_i$ is the dimensionality-reduced position of best matching unit (BMU) of $r_i$ within the trained SOM.
%
%Designed to achieve these qualities, the workflow of \ClimateSOM is shown in~\cref{fig:workflow} and is composed of the following steps:
\subsection{Workflow Overview}\label{ssec:workflow_overview}
\begin{figure*}[h]
    \centering
    \includegraphics[width=0.9\textwidth]{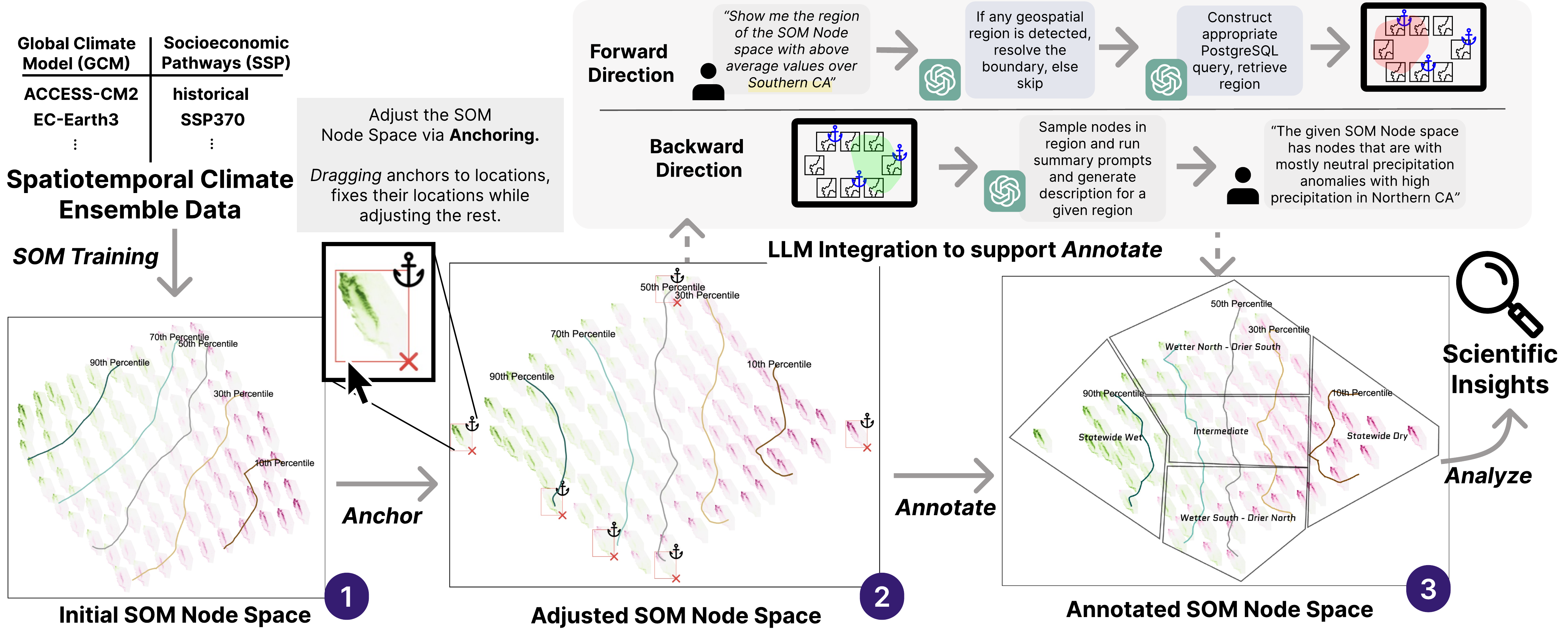}
    \caption{The \ClimateSOM workflow. The workflow is composed of the SOM Training step and three main steps: Anchor, Annotate, and Analyze. In \step{S1}: Anchor, the user adjusts the Initial SOM Node Space \subimage{1} using iterative anchoring with Minimum-Distortion Embedding to produce the \textit{Adjusted SOM Node Space} \subimage{2}. Dragged nodes are called \textit{anchors} and are fixed in place, while the rest of the nodes are adjusted. In \step{S2}: Annotate, the user annotates the SOM Node Space using manual inspection of the SOM Nodes or with the LLM Integration. Once annotated, in \step{S3} the user can
        use the \textit{Annotated SOM Node Space} \subimage{3}
    as the basis to analyze the ensemble dataset.}
    \vspace{-0.3in}
    \label{fig:workflow}
\end{figure*}

The workflow contains one pre-step and three main steps (\cref{fig:workflow}).
\begin{itemize}[left=0pt, itemsep=0pt, labelsep=3pt, align=left,  itemindent=0pt]
    \item[\step{S0} \textbf{Training the SOM}]
        The workflow begins with training the SOM on the dataset to ensure it has sufficient explanatory power---specifically, that the approximation via its BMU is adequate---and smoothness across the SOM grid.
        % A crucial assumption of \ClimateSOM is that SOM must capture \textit{sufficient} variance in the ensemble data, so that $r_i \approx BMU(r_i)$ is a reasonable approximation.
        % Careful consideration is required for the parameters for the SOM training and the metrics with which to evaluate and these are described in~\cref{ssec:som_training_parameters,ssec:som_training_metrics}.
        % The SOM is trained to capture the spatial variability of the input data set. The SOM is trained using the \textit{Kohonen} algorithm~\cite{kohonen1982self}, which is a type of unsupervised learning algorithm that is used to produce a low-dimensional representation of the input data set.
    \item[\step{S1} \textbf{Anchor}]
        Using the pseudo-2D smooth SOM nodes from Step 0, this step enables users to create the \textit{Adjusted SOM Node Space} by projecting the SOM nodes into a 2D layout. To support interpretation and navigation, users can anchor specific nodes---fixing their positions while allowing others to adjust---resulting in a more interpretable spatial arrangement than the original SOM grid. 
        We refer to this refined space as the \textit{Adjusted SOM Node Space}.
        %
        % Here, instead of using common techniques like t-SNE\cite{van2008visualizing} or UMAP\cite{mcinnes2018umap}, we opt for Minimum-Distortion Embedding (MDE) \cite{agrawal2021minimum}, which generalizes t-SNE and UMAP and crucially allows the anchoring of points in the output 2D space.
        % %
        % When supplied with a anchoring constraint, MDE ensures that these points are \textit{fixed} in a given location in the output space, while reducing the distortion of the remaining points.
        % %
        % We prefer MDE over t-SNE and UMAP since often fail to preserve important characteristics of the embedded space like interpretable axes.
        %
        % We use MDE to allows users to iteratively construct the \textit{Adjusted SOM Node Space}, anchoring points within the embedded 2D space to ensure a more meaningful layout of the SOM nodes.
    \item[\step{S2} \textbf{Annotate}]
        Once the \textit{Adjusted SOM Node} Space is defined, users annotate it to produce the \textit{Annotated SOM Node Space}. An annotation consists of (1) a bounded region within the adjusted space and (2) a text label describing that region. These annotations serve as an abstraction layer to facilitate interpretation in the next step. \ClimateSOM incorporates LLMs to assist with this process.
    \item[\step{S3} \textbf{Analyze}]
        A given model run can now be described as a distribution within the \textit{Annotated SOM Node Space}\revdel{, as well as}\revadd{. This distribution can be further broken down by user-defined \textit{annotations}}\revdel{ percentages broken down by user-defined \textit{annotations},} \revadd{by considering the number of time steps of that fall within each \textit{annotation} region} \req{R1}. Model runs can be compared either by directly comparing their distributions or by interpreting them as vector fields over the annotated space \req{R2}. Clustering can be applied to model runs based on these distributions, and to SSP forcings over each GCM by clustering changes in their respective distributions \req{R3}. \revdel{Please see the details}\revadd{Details are} in~\cref{sec:workflow}.
\end{itemize}

\noindent
Having outlined the workflow, we now address two important design choices: the use of SOMs and the intentional abstraction away from temporal ordering.

\vspace{3pt}
\noindent\textbf{SOM as the central tool.} 
We choose SOMs over other DR or clustering methods for the following two reasons:
\begin{enumerate}[noitemsep, topsep=0pt,leftmargin=12pt]
\item \textbf{Preservation of local topology}: SOMs preserve neighborhood relationships in lower-dimensions which enables meaningful spatial interpretations. Our distributional abstraction of model runs requires that the learned grid be continuous and neighboring nodes express similar spatial patterns. Other DR or clustering methods do not produce topologically smooth results in lower-dimensions and thus do not satisfy the conditions required in \ClimateSOM.
\item \textbf{Established use in climate science}: SOMs are widely used for analysis of climate data and are familiar to domain experts, aiding in its interpretation.
% \item \textbf{Visual continuity and smoothness}: Our data distributional abstraction of model runs requires that the learned grid must be continuous and neigh
\end{enumerate}

% Among the various possible methods to initially reduce the dimensionality of the input ensemble, \ClimateSOM opts for SOM for two main reasons: (1) SOMs are an established tool in climate science and are known to be successful in extracting salient features from spatiotemporal climate data and (2) \ClimateSOM relies on preserving the variability of the input ensemble via a pseudo-smooth 2D representation.  
%
% Maintaining smoothness in this representation is crucial, as discontinuities or irregularities can compromise the interpretability of the abstraction.
%
% While smoothness can be defined in various ways, we require that, locally at each node, the neighborhood maintains a consistent level of smoothness across the grid. 
%
% This ensures that the MDE embedding, powered by local neighborhood graphs, produces interpretable axes in 2D.
% \note{How do I properly define meaningful after MDE?}
%
% SOMs with the tuned parameters can guarantee this; however, other methods like hierarchical or \textit{k}-means clustering do not.

\vspace{3pt}
\noindent\textbf{Ignoring the temporal ordering.}
\ClimateSOM represents a model run $R$ as a set of points $S$, thus discarding temporal order (scrambling R yields the same S). This is an intentional trade-off for two reasons:
\begin{enumerate}[noitemsep, topsep=0pt,  leftmargin=12pt]
    \item \textbf{Emphasis of broad patterns}: The goal is to analyze the general behavior of model runs over time rather than their specific sequence. This aligns with how GCM outputs are studied for broad differences in climate science.
    \item \textbf{Reducing complexity}: While preserving temporal order enables the use of time series analysis techniques, such as dynamic time warping, it also introduces complexity and amplifies the random variations inherent in climate ensemble data.
\end{enumerate}
This trade-off enables simpler interpretation while retaining the ability to extract broad patterns from model outputs.
% %
% (1) \
% %
% (2) while accounting for the temporal order allows for the application of time series  analysis and techniques like dynamic time warping, it is certain that it also introduces complexity and accentuates the random variations that climate ensemble data exhibits.
%
% Despite the loss of the temporal ordering, we argue that \ClimateSOM can still produce meaningful insights into the behavior of climate ensemble datasets.

%% file: tex/4_workflow.tex
% We describe the steps of the \ClimateSOM workflow in detail in this section.
\subsection{Training the SOM (\step{S0})}
% \subsection{Self-Organizing Maps (SOM)}
% Self-organizing maps are a type of neural network designed to learn the topology of the input data by mapping it to a lower-dimensional grid, usually in a rectangular grid in 2D of size $dim\times dim$.
%
% We refer to \cite{kohonen_self-organizing_1990} for its formal definition.
%
% After a random initialization, at each iteration, the SOM algorithm selects a random input field and finds the best matching node (BMU) in the grid. 
%
% While varying definitions of \textit{best matching} exist, we use the Euclidean distance between the input field and the node's field as the distance metric.
%
% After identifying the BMU at each training step, the algorithm updates the SOM weights based on a learning rate and a neighborhood function.
%
% Formally, for a data set $R=\{r_1, r_2, \dots r_n\}$ of a collection of spatiotemporal fields (in the case studies, the fields are a series of precipitation anomalies over chosen domains in the Western US), at every iteration $t$, the algorithm selects a random sample $x(t)$ from $R$ finds its BMU, node $b$, and the weight vector $w_i$ of the node $i$ is updated as follows:
% \begin{equation}\label{eq:som_update_rule}
%     w_{i}(t+1) = w_{i}(t) + \alpha(t)h_{i,b}(t)[x(t) - w_{i}(t)]
% \end{equation}
% where $w_{i}$ is the weight vector of the node $i$, $\alpha(t)$ is the learning rate at time $t$, $h_{i,b}(t)$ is the neighborhood function that computes the influence of node $i$'s update, defined as a function of the distance between nodes $i$ and $b$.
%
For our spatiotemporal ensemble dataset, we train the SOM on the flattened ensemble dataset (i.e. across all spatial realizations that occur in the ensemble).
As common choice, we opt for the neighborhood function $h(t)$ to be the Gaussian function and linearly decrease its $\sigma$ as is standard practice in climate applications~\cite{gibson_use_2017}.
After training, the grid of SOM nodes is expected to approximate and learn the distribution input ensemble dataset under the constraint that the 2D grid of nodes are pseudo-smooth.

The SOM must meet two competing criteria: (1) it should retain sufficient information from the input data, and (2) the 2D grid of nodes should be smooth enough to avoid disconnections after projection, which would hinder interpretability.
However, these two criteria are often in conflict. 
%
% Smaller neighborhood update size often results in higher explanatory power allowing the SOM to extract finer details of the input data's topology; however, this is at the cost of non-smooth SOM grid since each update influences a smaller region of the grid, and vice versa.
%
% A smaller neighborhood update size enhances the SOM’s ability to capture finer details of the input data’s topology, increasing its explanatory power. 
%
% However, this comes at the cost of a less smooth SOM grid, as each update affects a smaller region. 
%
% Conversely, a larger neighborhood size promotes smoother transitions but may obscure finer details.
%
A smaller neighborhood update size captures finer details of the data’s topology but results in a less smooth SOM grid, as each update affects a smaller region. 
Conversely, a larger update size smooths the grid but may obscure finer details.
We found that the choice of (1) the initial neighborhood update size ratio expressed as a ratio between the initial update size and the dimension of the SOM grid, ($kR=(\sigma_{initial})^2/ {dim}^2$), and (2) the ratio of the final neighborhood update size, $kS=\sigma_{final}/ \sigma_{initial}$, to the initial size at the first iteration ($kS$) are critical to balancing these two criteria.
For our case studies, we found that $kR=0.03$ and $kS=0.2$ provide balanced results.
Further details are in the Appendix.

\subsection{Anchor (\step{S1})}
% \subsubsection{Dimensionality Reduction (DR) on the SOM Nodes}
As seen in past works in climate science employing SOMs \cite{sheridan_self-organizing_2011}, the SOM grid, while constructed to be locally smooth due to the training process, can have distortions globally. 
Therefore, SOMs are often presented alongside Sammon maps\cite{sammon1969nonlinear} or Multidimensional Scaling (MDS) to provide a more faithful view of the SOM grid in terms of its relative distances between nodes.
However, these methods give users no control of the mapping process and gives priority to global structure in place of local structure.
To provide users with the iterative ability to shape and project the SOM nodes, we allow user to \textit{anchors} particular SOM node in the projection process.
% 
% In line with these approaches, we propose to apply a dimensionality reduction (DR) technique on the SOM grid nodes to yield a more interpretable 2D representation of the spatiotemporal fields in the input data set.
%
% However unlike similar approaches that rely of global DR methods like Sammon mapping or MDS, we propose a local DR method for two reasons. 
%
% First, the most interesting comparison arises between ensemble members whose variations are often local in nature. 
%
% First, the differences between ensemble members can be subtle since they all model similar phenomena and subtle differences are better captured by local DR methods.
%
% In downstream tasks like clustering or the characterization of forcing, we rely on collection of local differences to encapsulate overall differences, and thus preserving local structures is preferred over its counterpart.
%
% Second, Euclidean distance, our distance metric to define pairwise difference between SOM nodes, is a sufficient local metric but a poor global metric, and thus optimizing for the global structure may introduce distortions that are less desirable.
%
% An important component of this step is the anchoring of points in the output 2D space as a constraint to the DR process.
%
% This allows for the user to configure the space in which the ensemble data will be visualized by explicitly soliciting the user to reorganize the 2D space to better suit their needs.
%
The utility of placing constraints on DR processes has been explored in previous works \cite{vu_integrating_2022}, and we find the iterative nature of anchoring to be most suitable as it requires minimal intervention from the user.

%
% We claim that this is preferable over methods like UMAP as these methods can fail to preseve important characteristics of the embedded low-dimensional space, such as the smooth contours of the SOM grid.
%
In \ClimateSOM, we employ Minimum-Distortion Embedding (MDE)\cite{agrawal2021minimum} to support anchoring.
Similar to UMAP\cite{mcinnes2018umap}, MDE can be presented as an optimization problem over local neighborhood graphs.
% , where the distortion between any two input data $i$ and $j$ arising from the dimensionality reduction is defined as a function of $d_{i,j}$, the Euclidean distance between the $x_i$ and $x_j$ -- embeddings of $i$ and $j$. 
%
%This function is formalized by Agrawal et al. as $f_{(i,j)}(d_{i,j})$ and referred to as the \textit{distortion function}.
%%
% Letting $\varepsilon$ to be the set of all pairs of $(i,j)$ and defining the \textit{average distortion} as $E(X)=(\Sigma_{(i,j)\in\varepsilon}f(i,j)) / |\varepsilon|$, MDE is formulated as the optimization problem to minimize $E(X)$ over all sets of feasible embeddings of the input $X$.
%
% Depending on MDE's configuration, this generalization can represent various DR techniques from PCA, MDS to UMAP.
%
Using MDE, we optimize for the distortion of the SOM grid nodes in the 2D space while anchoring points in the output space to ensure a more meaningful layout of the SOM nodes.
\revdel{Since a 2D SOM roughly extracts two main axes of variation, this step allows user to physically map these axes to the two axes of the visual interface itself.}
\revadd{In our UI, dragging the SOM Nodes designates them as \textit{anchored} and are fixed in place in the projection, while other nodes are adjusted.}
\revadd{This anchor step allows users to (1) physically map the two main axes of variation that the 2D SOM extracts to the visual interface and (2) semantically map nodes to appropriate locations in the screen. (e.g. SOM nodes with high value in the \textit{north} be placed in the \textit{upper-half} of the screen.)
We find that fine adjustments like encouraging specific nodes to be positioned at particular locations required iterative methods. This anchoring interaction was well suited, and thus included in \ClimateSOM.}
We refer to the output of this step as the \textit{Adjusted SOM Node Space}.

\subsection{Annotate (\step{S2})}\label{ssec:annotate}
This step allows users to provide \textit{annotations} to the \textit{Adjusted SOM Node Space} to add user-defined context to the SOM nodes in preparation for subsequent analysis in \step{S3}.
We define an \textit{annotation} as a pair of the following: (1) a bounded region within the \textit{Adjusted SOM Node Space} and (2) a text label that describes the region.
Two challenges arise: identifying relevant regions and defining meaningful labels.
Given the documented success of Text-to-SQL via LLMs \cite{zhang_natural_2024} and text summarization \cite{pu_summarization_2023}, our approach for \ClimateSOM is leveraging LLMs in this process, alongside the user's visual inspection of the individual SOM nodes.
These correspond to the forward and backward directions of LLM assistance as shown in \cref{fig:workflow}.

\vspace{-3pt}
\subsubsection{Forward Direction}
In the forward direction, we assume the user has a characteristic of interest in mind for SOM nodes (e.g., ``\textit{Show me SOM nodes with average precipitation over Southern California above 0}''). The goal is to identify the corresponding region in the \textit{Adjusted SOM Node Space} by filtering nodes that match the specified condition. This is done by generating a PostGIS query based on the user’s geospatial query.
However, we found that naively prompting this task to the LLM does not yield consistent results, primarily due to LLM's
% An identified challenge with tasking LLMs to generate spatial queries from natural language, however, is the 
lack of accuracy in generating appropriate boundaries for given geospatial regions as found in \cite{liu_nalspatial_2025}.
To address this, we delegate the LLM not to generate spatial boundaries, but instead to generate a list of counties that correspond to the named region. These counties, which have well-defined boundaries, are then used to construct the spatial filter, reducing errors introduced by the LLM.
%
% In addition, we also adopt self-correction for the SQL query generation as seen in \cite{pourreza_din-sql_2023}.
%
In all, the forward direction proceeds as follows.
\begin{enumerate}[nolistsep,  leftmargin=12pt]
    \item Parse the user query to detect any region names or locations. If none are detected, skip to step 3.
    \item Resolve the identified regions to a list of counties with known boundary definitions. Combine these county boundaries to form the target region.
    \item Use the resolved region and the numerical conditions extracted from the query to construct and execute a PostGIS query over the SOM nodes.
\end{enumerate}
\paragraph{Supported Question Types}
Based on discussions with the domain experts, we identify three categories of user queries that the LLM should be able to support by generating appropriate PostGIS queries:
\begin{enumerate}[nolistsep,  leftmargin=12pt]
    \item SOM nodes such that the average value over \textit{some geospatial region} is above / below \textit{X}
    \item SOM nodes such that the average value over \textit{some geospatial region} is between \textit{X} and \textit{Y}
    \item SOM nodes such that the average value over \textit{some geospatial region A} is higher or lower than the average value over \textit{some geospatial region B}
\end{enumerate}

\begin{figure*}[!t]
    \centering
    \includegraphics[width=0.95\textwidth]{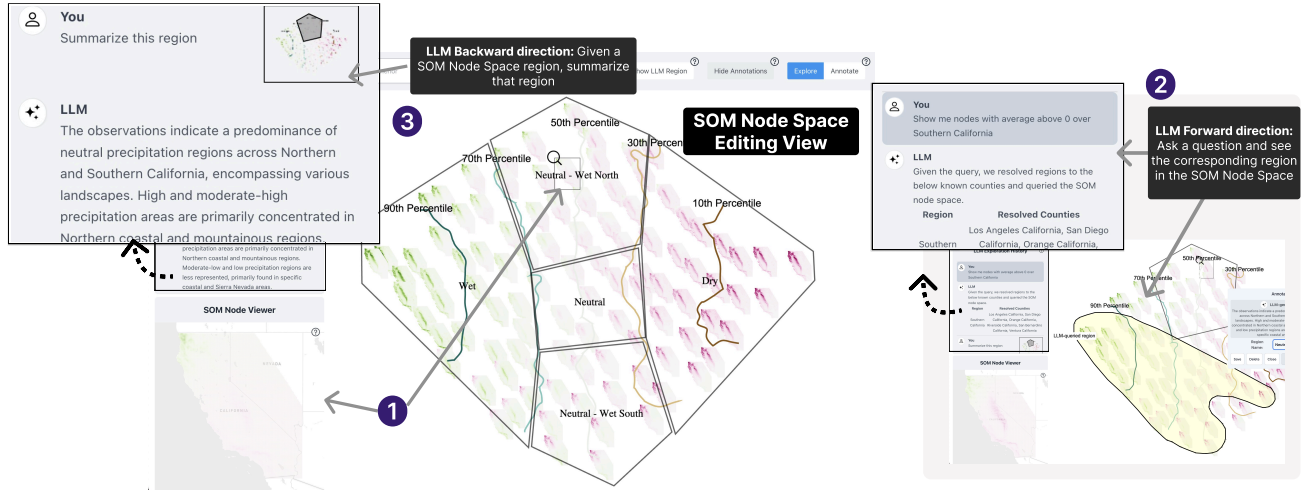}
    \caption{
        The \ClimateSOM interface at the end of \step{S1} and \step{S2} displaying the \textit{Annotated SOM Node Space}.
        %
%        The visual interface to support the first two steps --- \step{S1}: Anchor and \step{S2}: Annotate. 
%        %
%        The assumed user is first presented with \highlight{1}. 
%        %
        The user can inspect each SOM node via clicking on the node \subimage{1}, and the Node Viewer on the left displays the selected map of the SOM Node.
        %
        % After some initial visual inspection, they realize that the initial embedding of the SOM Nodes are not optimal since they cannot well interpret the dimensionality reduced axes. 
        %
        % The user then drags SOM Nodes around \subimage{1b} and clicks \textit{Reflect Anchors} and iteratively arrives to \highlight{2}. 
        %
        To support \step{S2}: Annotate, users can leverage the two directions of LLM-powered query system: Forward \subimage{2} and Backward \subimage{3}. 
        For the forward direction \subimage{2}, the user can request a particular quality of interest, (e.g., \textit{Show me nodes that have average above 0 over Southern California}) and be presented with the corresponding SOM Node region bounded in yellow. 
        For the backward direction \subimage{3}, the user can \revdel{select}\revadd{right click on} a SOM Node region and a LLM will generate a \textasciitilde
 50-word summary of SOM Node realizations in that region.
        %
        % The user's interaction with the LLM are tracked in \subimage{2d}.
        %
        Through visual inspection and the LLM assistance, the user compiles the set of \textit{annotations} as regions in the SOM Node Space and a label corresponding to the region.}
        \vspace{-0.3in}
    \label{fig:Step1and2}
\end{figure*}
\vspace{-3pt}
\subsubsection{Backward Direction}
In the backward direction, we presume the user has identified a region within the \textit{Adjusted SOM Node Space} and wishes to get a high-level summary of the region. 
In \ClimateSOM we accomplish this as follows.
\begin{enumerate}[nolistsep,  leftmargin=12pt]
    \item Given a region in the \textit{Adjusted SOM Node Space}, we extract samples evenly within the SOM Node Space region. 
        % For our case studies, we found nine samples to yield sufficient results.
    \item Construct a per-county level summary statistics of each node from step 1. For a given sampled SOM node, we textually represent \revdel{each node} as: 
    \vspace{-4pt} 
%     \revdel{\begin{verbatim} 
% { high_value:            [counties_list], 
%   moderately_high_value: [counties_list],
%   neutral_value:         [counties_list],
%   moderately_low_value:  [counties_list],
%   low_value:             [counties_list] }
% \end{verbatim} }
\begin{verbatim}
  { high:[counties_list], mod_high:[counties_list], 
  neutral:[counties_list], mod_low:[counties_list], 
  low:[counties_list] }  
\end{verbatim}

% \begin{lstlisting}[xleftmargin=0pt, frame=none, basicstyle=\small\ttfamily]
% { high: [counties_list], mod_high: [counties_list], 
%   neutral: [counties_list], mod_low: [counties_list], 
%   low: [counties_list] }  
% \end{lstlisting}
\vspace{-4pt}
    That is, we enumerate all counties with average values in each of the five categories.
    Note that, we allow \revdel{the} the cutoff values that define \textit{high}, \textit{moderately high}, etc. to be modified by the user.
    \item Use the text-based summary statistics for each sampled SOM node with the LLM to generate a summary description of the region.
\end{enumerate}

\vspace{3pt}
\noindent
In summary, the LLM is used in the forward direction for two tasks: (1) mapping region names to a list of counties, and (2) generating PostGIS queries to retrieve relevant SOM nodes. In the backward direction, the LLM serves as a summarizer of per-county statistics derived from user-selected regions.
We implement both directions using LangChain \cite{langchainLangChain} with OpenAI's \textit{gpt-4o-mini}\cite{achiam2023gpt} and use LangChain Tools for the forward direction to detect if the user input includes any geospatial regions that need to be resolved.
Details of the prompts used are available in the Appendix.
Either via manual inspection of the \textit{Adjusted SOM Node Space} or via the LLM integration, the user can now define a set of \textit{annotations} for the \textit{Adjusted SOM Node Space} to produce the \textit{Annotated SOM Node Space}.

\vspace{-3pt}
\subsection{Analyze (\step{S3})}
Finally in this step, we describe how \ClimateSOM performs each of the tasks and meets the design requirements outlined in~\cref{ssec:design_requirements}.
\vspace{-3pt}
\subsubsection{Exploring a single set of model runs \req{R1}}\label{ssec:explore}
% \begin{figure}[htbp]
%     \centering
%      \includegraphics[width=\linewidth]{figures/ExploreOne.pdf}
%      \caption{}
% \end{figure}
Given a single ensemble model run with $n$ time steps, $R=[r_1, r_2,\ldots, r_n]$, we can describe it as a distribution $S=\{p_1,p_2,\ldots, p_n\}$ within the \textit{Annotated SOM Node Space} by mapping each $r_i$ to its best BMU and applying the MDE projection: $p_i = MDE(BMU(r_i))$.
We then compute a kernel density estimate (KDE) over $S$, which serves as a “fingerprint” of the model run. Regions of high density indicate spatial patterns that occur frequently in the run \revadd{(\cref{fig:Step3}-\subimage{1})}. This distribution can also be expressed as percentages over the annotated regions defined in \step{S2}, enabling semantic interpretation beyond the KDE plot  \revadd{(\cref{fig:Step3}-\subimage{2})}.
For a set of multiple model runs, we apply the same process over the entire set and consider the \revdel{aggregated set of}\revadd{set of all} points as the representative distribution.
\subsubsection{Comparing the behavior of two sets of model runs \req{R2}}
% \begin{figure}[htbp]
%     \centering
%      \includegraphics[width=\linewidth]{figures/Compare.pdf}
%      \caption{}
% \end{figure}
To facilitate comparison of two model runs $R_1$ and $R_2$, \ClimateSOM provides two methods: (1) side-by-side comparison and (2) a vector field formulation.
\paragraph{Comparison via side-by-side}
The side-by-side comparison is a direct comparison of the distributions of $R_1$ and $R_2$ in the \textit{Annotated SOM Node Space} as defined in \cref{ssec:explore}.
By presenting these two views side-by-side, we allow users to visually inspect if there are regions within the SOM Node Space that are occupied in similar densities or different densities by $R_1$ and $R_2$ \revadd{(\cref{fig:sidebyside}}).
Similar to the exploration of a single model run, the respective distributions can be compared via the breakdown based on the annotations defined in \step{S2} Annotate. 

\paragraph{Comparison via a vector field formulation}
We also support a formal comparison using vector fields, based on a bootstrap approach \cite{efron1992bootstrap} (implementation details in the Appendix).
Given two model runs $R_1$ and $R_2$, we generate $k$ number of bootstrap samples for each model run.
At each iteration, we first calculate the 2D optimal transport\cite{flamary2021pot} between the two bootstrap samples.
Given the optimal transport plan, we construct a set of vectors for each pair in the transport plan that points from its source to its target.
% we consider each pair in the plan and construct a set of vectors that points from each source to its target.
%
Interpolating the results for the set of vectors onto a regular grid generates a vector field, representing the spatial transition from $R_1$ to $R_2$ \revadd{(\cref{fig:Step3}-\subimage{3})}.
%%
%Letting $R_1',R_2'$ be a bootstrap samples, for each bootstrap iteration we first compute the 2D optimal transport\cite{flamary2021pot} between $R_1'$ and $R_2'$.
%%
%This results in a list of pairs of elements $(r_1',r_2')$ where $r_1'\in R_1'$ and $r_2'\in R_2'$ such that the total cost of transport or matching is minimized.
%%
%Using this 2D optimal transport result, we aggregate and linearly interpolate this result onto a regular grid to generate a vector field.
%%
%Finally, averaging this result from all bootstrap iterations, we can generate a vector field that represents the difference between $R_1$ and $R_2$ in the \textit{Annotated SOM Node Space} as a transition from $R_1$ and $R_2$.
%
% Thus, the critical assumption of this approach is that is that the \textit{change} from $R_1$ to $R_2$ can be estimated by the 2D optimal transport between the two distributions.
%
% For our case studies, we found $k=50$ to produce stable results.
%
The resulting field represents the spatial transition from $R_1$ to $R_2$.
This further enables analysis at the level of annotations---e.g., determining where points from annotation A in $R_1$ are mapped in $R_2$ \revadd{(\cref{fig:Characteristic})}.

This approach is particularly useful when comparing temporally ordered scenarios (e.g. comparing a historical run to a future run or comparing a low to a high emission scenario). The vector field comparison may provide an added layer of context since the comparison is treated as a \textit{transition} from one to the other.
When comparing between two sets of model runs, we apply the same process to the aggregated set of points in each set as was in \req{R1}.

% \begin{figure}[h]
%     \centering
%     \includegraphics[width=\linewidth]{figures/BootstrapComparison.pdf}
%     \caption{The bootstrap-based construction of comparison for \ClimateSOM as a vector field}
%     \label{fig:bootstrap_comparison}
% \end{figure}

\vspace{-3pt}
\subsubsection{Clustering the behavior of many model runs \req{R3} } \label{ssec:clustering}
% \setlength{\intextsep}{0pt} % Remove space above and below wrapfigure
% \setlength{\columnsep}{5pt} % Reduce space between text and figure
% \begin{wrapfigure}{o}{0.65\linewidth}
% \begin{figure}[htbp]
%     \centering
%      \includegraphics[width=0.65\linewidth]{figures/Cluster.pdf}
%      \caption{}
% \end{figure}
% \FloatBarrier % Forces wrapfigure to stay in place
% \setlength{\intextsep}{12pt} % Remove space above and below wrapfigure
% \setlength{\columnsep}{10pt} % Reduce space between text and figure
% The third and final task is to cluster the behavior of many model runs based on their distributions in the \textit{Annotated SOM Node Space}. 
%
% For clustering, we consider two variants.
\revdel{ClimateSOM supports two types of clustering as follows.}
\paragraph{Clustering model runs directly} Given the distributions of the model runs in the \textit{Annotated SOM Node Space}, we perform clustering using Earth Mover's Distance (EMD)\cite{flamary2021pot} to quantify the similarity producing a distance matrix between the distributions.
We then apply UMAP on the distance matrix to produce a 2D embedding and HDBSCAN \cite{campello2013density} to perform the clustering.
While HDBSCAN can operate on the distance matrix directly, we found more consistent results using this two-step method.
This results in a clustering of model runs that exhibit \textit{similar} and \textit{dissimilar} behavior, where each cluster can be represented as an aggregate distribution of the model runs in that cluster.
\paragraph{Clustering GCMs based on their SSP forcing} Given vector field formulation of forcing (i.e., the effect of an emission future as a \textit{transition} from the historical time period), we perform clustering of the vector field using an average cosine similarity to quantify the similarity between the vector fields.
Similarly, we can use UMAP and HDBSCAN to perform clustering such that we can identify GCMs that exhibit \textit{similar} and \textit{dissimilar} behavior under the same SSP forcing.
For this variant, each cluster can be represented as a vector field that represents the average forcing of the GCMs in that cluster.

A notable design implication of \ClimateSOM and its formulation is that the resulting clustering is interpretable, since it can be displayed directly in the same \textit{Annotated SOM Node Space} as distributions or vector fields.
%
% Within the broader domain of clustering algorithms and Explainable AI, there remains significant interest in the interpretability of various clustering algorithms \fix{cite}, as complex clustering workflows can often resemble a black box, making it difficult to discern the decision boundaries between clusters.
%
We consider this explainable clustering results as a key strength of \ClimateSOM\,\revadd{(\cref{fig:Step3}-\subimage{4})}.
%
% In contrast, the clustering for \ClimateSOM can be visualized directly in the \textit{Annotated SOM Node Space} as distributions or vector fields, providing a clear explanation of the clustering results.

% the SOM grid can be used as a lower-dimensional representation of the input data set.
% Once trained, the SOM  

% \subsection{Enabling interpretability of projected SOM node space}
% \subsection{Spatiotemporal data as a distribution}
% Q: Why not use DTW or other time-series distance metric?\\
% A: There are applications like some climate scenarios, where there are meaningful interpretations of the time series discarding the particular order, but considering the time series as a whole. Of course, we lose the temporal information, but we gain in simplifying the comparison.

% \subsection{Downstream tasks: Clustering and Member Comparisons}

%% file: tex/5_interface.tex
To introduce the interface design for \ClimateSOM, let's consider a climate scientist with a climate ensemble dataset of monthly precipitation projections over California, expressed an anomalies from historical averages.
Given this ensemble, we assume the scientist in interested in the three research questions introduced in \cref{sec:problem}.

Assuming \step{S0} is complete, the trained SOM is a 2D grid of SOM nodes, where each node is a precipitation map over California.
Then the scientist proceeds through \step{S1} and \step{S2}.
% \subsection{\step{S1} Step 1: Anchor and \step{S2} Step 2: Annotate}

\subsection{SOM Node Space Editing View}
% \paragraph{Step 1: Anchor}
The main visual component for \ClimateSOM is the trained SOM and the SOM Node Space that the user iteratively edits as seen in~\cref{fig:workflow}.
The SOM Node Space Editing View supports the anchoring and annotating of the SOM Nodes in \step{S1} and \step{S2}.
\revadd{For our case studies, displaying all SOM Nodes risks unnecessary clutter. As such a subset, 100 in the case studies, are rendered. This was determined via trial and error, optimizing for minimal occlusion whist maintaining enough nodes to communicate the node structure.}
First, as in~\cref{fig:workflow}-\subimage{1}, the \textit{Initial SOM Node Space} is presented to the user using the unanchored MDE result, with each node depicting a precipitation map over California using a Green-Pink (high-low precipitation) colormap.
The interface also presents colored percentile lines in the SOM Node Space to orient the user to the mean anomaly in each SOM node.

% \begin{figure}[h]
%     \centering
%     \includegraphics[width=\linewidth]{figures/temp2.pdf}
%     \caption{Anchoring nodes reshapes the SOM Node Space in \step{S1}.}
%     \label{fig:step1}
% \end{figure}
% \label{sec:Step1and2Inteface}
% \cref{fig:Step1and2} provides an overview of the visual interface support for the two steps.
%
% First, the user is presented with \highlight{1} displaying the initial unanchored MDE result.
%
In \step{S1} the user constructs the \textit{Adjusted SOM Node Space} by anchoring and reflecting the anchors via MDE. 
SOM nodes in this step can be dragged around freely, and nodes that are set to be anchored are displayed with an anchor icon as seen in~\cref{fig:workflow}.
Furthermore, they can be inspected further by clicking on the node, displaying that node in the SOM Node Viewer (\cref{fig:Step1and2}-\subimage{1}).
Once they have a satisfactory \textit{Adjusted SOM Node Space} they proceed to \step{S2}.
\revadd{Right clicking within the \textit{Adjusted SOM Node Space}, users can define vertices for each \textit{annotation} polygon.}
This step constructs the \textit{Annotated SOM Node Space} by leveraging the LLM-empowered bidirectional query system in addition to manual visual inspection.
The interface supports two directions of LLM integration, forward (\cref{fig:Step1and2}-\subimage{2}) and backward (\cref{fig:Step1and2}-\subimage{3}), to assist the user in defining annotations.
The interface provides the history of the user's interaction with the LLM in a chat-like visual UI.
Annotations as seen in~\cref{fig:Step1and2} are rendered as bounded regions in the \textit{Annotated SOM Node Space} with a user-defined label.
This label can be produced via the user's inspection of the bounded region in question, or with assistance from the LLM's \textasciitilde
 50-word summary of the region.
Once all annotations are defined and named, the \textit{Annotated SOM Node Space} is ready for analysis in \step{S3}.
\cref{fig:Step3} provides an overview of the visual interface support for \step{S3}.
\subsection{Model Run Inspection and Characteristic View}
\begin{figure}[htbp]
    \centering
    \includegraphics[width=\linewidth]{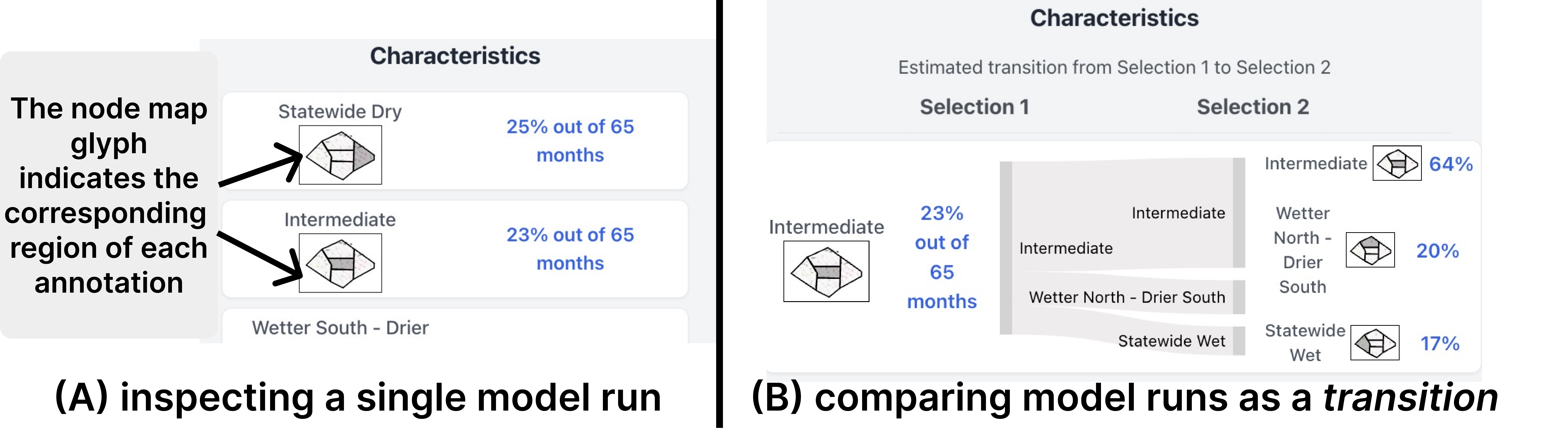}
    \caption{The Characteristic Viewer for inspecting single model run and when comparing two model runs as a vector field.}
    \vspace{-0.1in}
    \label{fig:Characteristic}
\end{figure}

% \paragraph{Model Run Inspection \subimage{1} and Characteristic View \subimage{2}}
First in pursuit of \req{R1}, the user can inspect a single model run over any given month by selecting them via a dropdown menu and view its distribution over the \textit{Annotated SOM Node Space} in the Model Run Inspection View (\cref{fig:Step3}-\subimage{1}).
The interface presents the 2D distributional representation of the selected model run as a KDE plot drawing three polygons as of the top 25\%, 50\%, and 75\% density regions.
Thus, the darkest areas of the KDE plot correspond to the SOM Node Space regions with the highest density of realizations in the selected model run.
In addition, we provide the breakdown of that model run's behavior broken down by the annotations in the Characteristic View (\cref{fig:Characteristic}).
In the Characteristic View, the distribution for the selected model run is broken down by the annotations, presented alongside a node map glyph (\cref{fig:Characteristic}-left), ordered by its percentages in each annotation region. 
The node map glyph is designed to quickly direct the user to the SOM Node Space region that each annotation represents.
Furthermore, in the case that the user is comparing two model runs as a vector field, the Characteristics View instead displays the \textit{transition} as a Sankey diagram between the annotations.

\noindent
\quad\textit{Justification:}\,\, In exploring the behavior of model runs, \textbf{the primary goal for climate scientists is to understand where the \textit{hotspots} are}.
That is, which part of the SOM Node Space most of the realizations of the model runs are concentrated. 
To this end, we employ the KDE representation as the visual encoding for the distribution so that we draw the most attention to the \textit{hotspots} in the distribution. 
Other designs we considered include using a 3D representation and encoding the density in the height, but we found a 3D encoding less effective as it was more difficult to interpret. 
The Characteristic View is designed in a similar vein, to provide a quick overview of the distribution of the model run over the annotations provided in \step{S2}, and the node map glyph directing the user to the \textit{hotspots} at the level of annotations in the distribution.

\subsection{Model Comparison View}
\begin{figure}[htbp]
    \centering
    \includegraphics[width=0.9\linewidth]{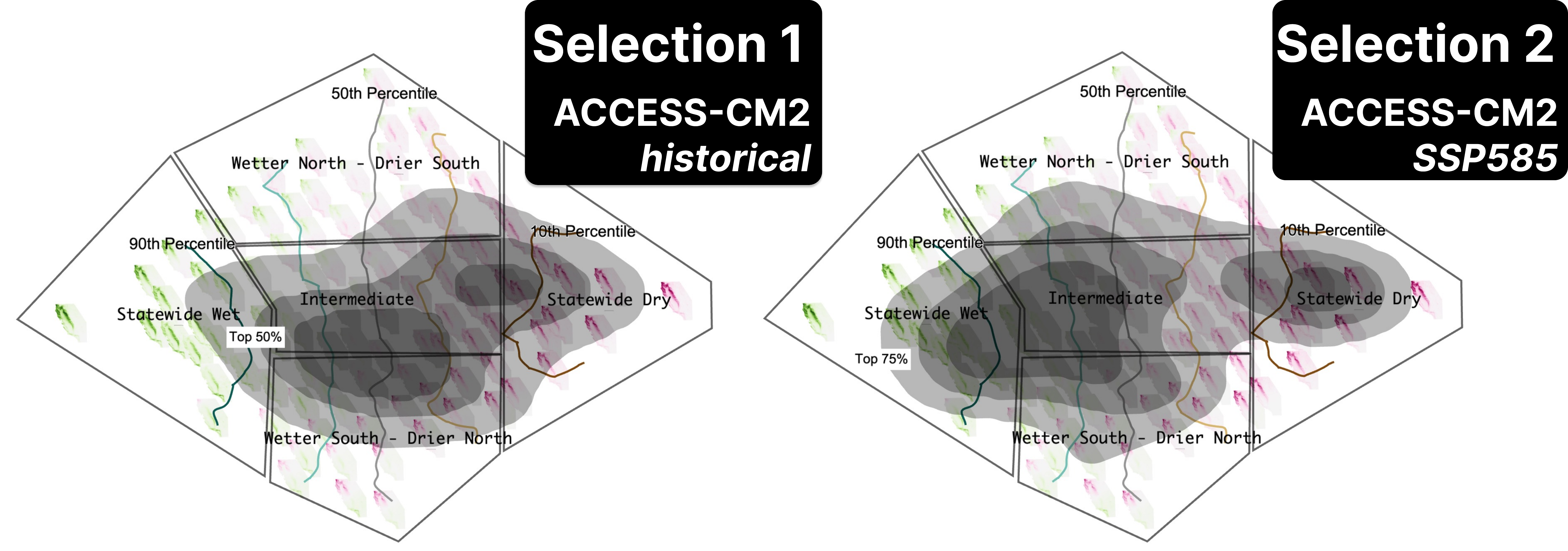}
    \caption{The side-by-side comparison view for comparing two model runs.}
    \vspace{-0.1in}
    \label{fig:sidebyside}
\end{figure}

% \paragraph{Comparison Views ~\cref{fig:sidebyside} and \cref{fig:Step3}-\subimage{3}} 
To support \req{R2}, we offer two visual interfaces for comparing two model runs: via a side-by-side comparison and as a vector field.
First, the side-by-side comparison allows for simple visual comparison of two model runs, by placing the two distributions side-by-side in the \textit{Annotated SOM Node Space} (\cref{fig:sidebyside}).
By drawing the distributions over the identical SOM Node Space, the user can quickly identify the SOM Node Space regions where the densities differ.
However, to support the user for comparative tasks where there is a natural order to the comparison (e.g., comparing the same GCM over the historical vs. future period), we provide a vector field comparison view as well(~\cref{fig:Step3}-\subimage{3}).

\quad\textit{Justification:}\,\, 
% When considering comparison of two model runs, over discussions with our domain expert, we identified the two types of comparisons, one where is a natural ordering to the comparison and one where there is not.
% Therefore, our visual interface was designed to support both types of comparisons. 
%
Although the side-by-side comparison is simple to interpret, there were cases where the differences were subtle, or when we could infer an ordering, there was scope to provide a simpler visual encoding for the comparison.
We found that the vector field representation was a good fit for this task as the direction of the arrows provided a natural interpretation of the comparison as a \textit{transition}.

For the vector field formulation of the comparison, we initially used a particle advection animation similar to those used in flow visualization but found that the static vector field representation to be more simple and required less visual processing.
In all, the two methods for comparison were well received by our domain expert, and thus implemented in the final design.

%In \subimage{3c}, the user can inspect the breakdown of the model run over time, here for January, across the annotations provided in \step{S2}.
%%
%Secondly, to support \req{R2}, the user can compare two model runs via two visual interfaces, side-by-side \highlight{2} and via a vector field \highlight{3}.
%%
%With reference to the vector field \highlight{3} the user can understand the \textit{transition} from Selection 1 to Selection 2 with reference to the Sankey-like diagram \subimage{3e} displaying the best estimate of this transition considering the \textit{destination} of points originally in each annotation region in Selection 1.
%
\subsection{Clustering View}
\begin{figure*}[!t]
    \centering
    \includegraphics[width=0.95\textwidth]{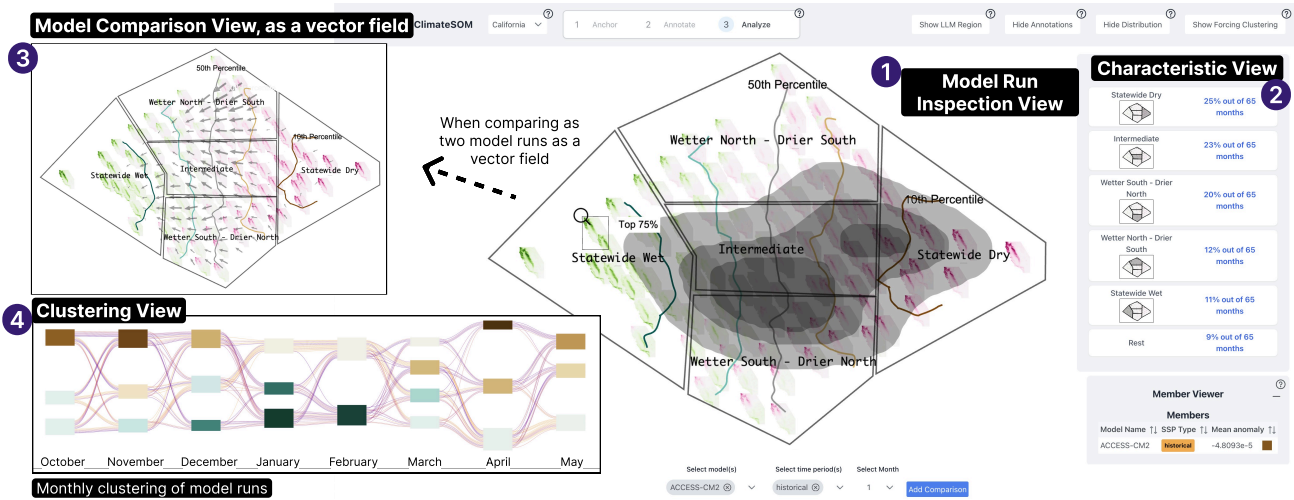}
    \caption{
        The visual interface to support the last step --- \step{S3} Analyze. 
        Each selected model run is displayed as a distribution over the \textit{Annotated SOM Node Space} \subimage{1} and the breakdown of the distribution over the annotations are displayed in the Characteristic View \subimage{2}.
        To support comparison, this main interface also supports a side-by-side comparison (\cref{fig:sidebyside}) but also as a vector field \subimage{3}.
        Lastly, to support clustering of model runs or GCMs by SSP forcing, the circuit-line diagram displays the monthly clustering results and their evolution \subimage{4}. 
        %%
        %First, in \highlight{1}, the user wishes to inspect a single model run: \req{R1}. 
        %%
        %The user can select a ensemble model run \subimage{3a} and view that model run as a distribution over the \textit{Annotated SOM Node Space} \subimage{3b}. 
        %%
        %Furthermore, given the annotations provided in \step{S2}, they are presented with a breakdown of the model run, here for January over time \subimage{3c}. 
        %%
        %The user can also inspect the list of the currently selected model \subimage{3d}. 
        %%
        %To support comparisons \req{R2}, we offer two visual interfaces, \highlight{2} and \highlight{3}, to support comparison as a side-by-side and via a vector field respectively. 
        %%
        %Of note is the comparison as a vector field in \highlight{3} where the user can inspect the comparison as a \textit{transition} from Selection 1 to Selection 2 \subimage{3e} and the interface provides a Sankey-like diagram to visualize out best estimate of transitions across the user-provided annotations. 
        %%
        %Finally \req{R3} is supported via a monthly clustering visualization \highlight{4}. 
        %%
        %\subimage{3f} and \subimage{3g} display the two variants of clusterings and the results as a circuit-line diagram connecting the per-month clustering results, placed according to 1D MDS results.
        %%
        %Interaction design and details are provided in \cref{fig:Step3Interaction} and \cref{sec:Step3Inteface}.
    }
    \vspace{-0.25in}
    \label{fig:Step3}
\end{figure*}

Lastly, to support \req{R3}, clustering results (\cref{ssec:clustering}) are visualized as a circuit-line diagram connecting the independent per-month clustering results (\cref{fig:Step3}-\subimage{4}).
For our scientists, \textbf{in addition to understanding the results of the per-month clustering results, there is also interest in understanding the evolution of the monthly clustering results.}
That is, how the cluster membership of a model run changes over different months.
%
% Therefore, in addition to visualizing the clusters, we also provide the circuit-line diagram to visualize the evolution as well.

\begin{figure}[htbp]
    \centering
     \includegraphics[width=\linewidth]{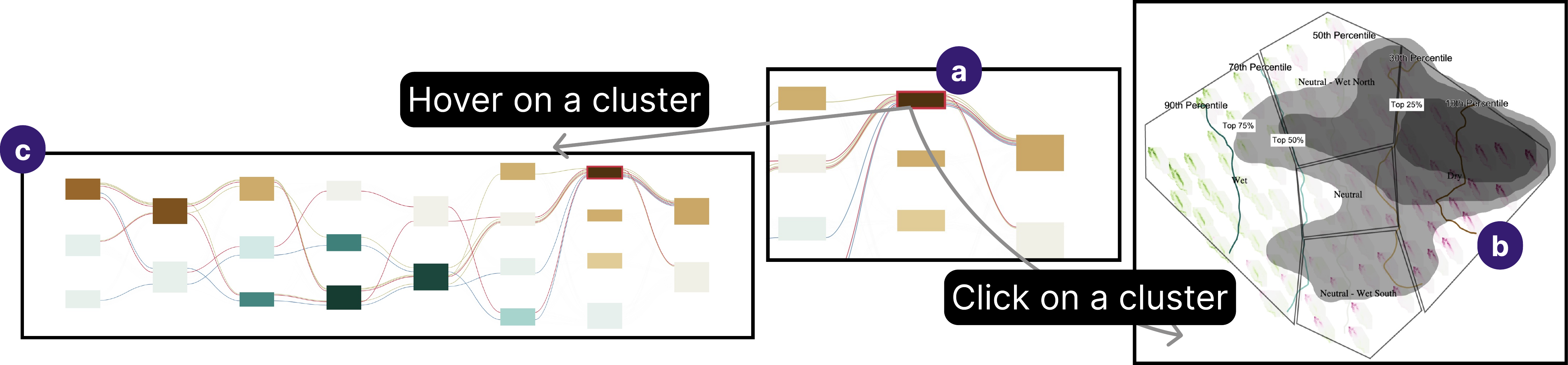}
     \caption{
     Clicking \subimage{a} on a monthly cluster model run clustering variant \cref{fig:Step3}--\subimage{4} will select model runs represented in that cluster \subimage{b} and show its distribution over the \textit{Annotated SOM Node Space}.
    On hover, only lines that pass through the hovered cluster are highlighted \subimage{c}.
    }
    \vspace{-0.1in}
    \label{fig:Step3Interaction}
\end{figure}

Each monthly cluster is represented as a box and the model runs are visualized as lines through the boxes, indicating its cluster membership for the given month.
The cluster boxes are colored according to the cluster's mean anomaly, and the lines according to the model run's SSP.
We compute a 1D MDS with intercluster distances as the distance matrix for each month, and the cluster boxes are placed according to this result.
Therefore, for every month, more similar clusters are placed closer together and less similar clusters further apart. 
In the clustering diagrams, tracing each line provides a sense of the evolution of the clustering results over the months.
%
% The lines can be filtered via the user's selection via \subimage{3h} and 
To support the investigation of the behavior of each cluster, clicking on a cluster will trigger the display of the data of that cluster as seen in \cref{fig:Step3Interaction}. 
Furthermore, to support the user in understanding if there are inter-month relationships between the clusters, we implement a hover interaction as well, by allowing the user to filter the lines based on those that pass through that hovered cluster.
A similar visual encoding supports the second variant of clustering--GCMs by forcing.
For this variant, each line represents GCMs instead of model runs as the GCMs are the subject of clustering.

% and we include the circuit-line diagram for this in the Appendix.

\noindent
\quad\textit{Justification:}\,\, As seen in works like \cite{jackle2016temporal} where the patterns in the data \textit{and} the evolution of patterns are the subject of inquiry, we apply a similar approach utilizing 1D MDS, letting the x-axis represent time and y-axis represent similarity.
Supporting the analysis of inter-month relationships in addition to the per-month clustering results was a key interest expressed by our domain expert, and thus we choose this design.

%% file: tex/6_usecases.tex
% \subsection{LOCA-Downscaled CMIP6 Precipitation Ensemble}

\subsection{Case Studies}
\paragraph{The Data}
To demonstrate that \ClimateSOM can uncover valuable insights, we apply the workflow to an ensemble dataset of LOCA-downscaled CMIP6 precipitation projections~\cite{pierce2014statistical,pierce2015improved,eyring2016overview} over California and the Northwest United States\cite{pierce2023future}.
%
% The Localized Constructed Analogs (LOCA) statistical downscaling method \cite{pierce2014statistical,pierce2015improved } is an analog-based method used to downscale
% relatively coarse scale global climate model projections from the CMIP6 archive \cite{eyring2016overview}  to finer scale regional projections. 
%
% The LOCA North American projections were carried out at 6km spatial resolution, in which downscaling of precipitation was conducted from global model projections made by 27 CMIP6 models, 
%
These model projections were made under 3 different SSPs \cite{riahi2017shared}, from a  SSP245, a relatively low greenhouse gas emission future, SSP370, a medium high greenhouse gas emissions future, and SSP585 a high greenhouse gas emissions future. 
%
% Of the 27 models, we consider \textasciitilde 14 models for each SSP and the historical period (i.e. four time periods per GCM) for which data was available, resulting in a total of 56 ensemble members.
We consider 14 GCMs for each SSP and the historical period (i.e. four time periods per GCM) for which data was available, resulting in a total of 56 ensemble members.
Details of the ensemble members used in this work can be found in the Appendix.
% LOCA or Localized Constructed Analogs, is a statistical downscaling method that takes coarse resolution climate model outputs, here of CMIP6\cite{eyring2016overview}, and downscales them to a regular grid of finer resolution.

We consider two regions separately: California and the Northwestern United States, encompassing Oregon, Washington, and Idaho.
While daily predictions are available, we focus on monthly projections considering months from October to May, as these months contribute to \textasciitilde85\% of the annual precipitation in both spatial regions.
\revdel{We apply two preprocessing steps.}
\revdel{Second, we}\revadd{In addition, we} normalize the monthly anomalies by dividing them by per-month standard deviation across models, time, and space.
The final data, therefore, consists of monthly precipitation anomalies expressed in standard deviations from the historical average.
\revadd{SOMs were trained on this monthly anomaly data across \textit{all} months, SSPs, and GCMs to best capture the overall anomaly structure.} 
For the California case study, each ensemble model run had a size of \revdel{$780\times 151\times 165$ (time $\times$ longitude $\times$ latitude), while for the Northwestern US, it was $780\times 60\times 115$} \revadd{$65\times12\times 151\times 165$ (years$\times$months$\times$longitude$\times$latitude), while for the Northwestern US, it was $65\times12\times 60\times 115$}.
We consult \textbf{E1}, a climate scientist and a co-author of this paper, for the California case study, and another domain expert \textbf{E2} for the Northwestern US case study.

% While this makes any analysis across different months difficult, this simplification allows us to train a single SOM across different months that streamlines the analysis.

\subsubsection{LOCA-Downscaled CMIP6 over California}
\begin{figure}[!t]
    \centering
     \includegraphics[width=0.9\linewidth]{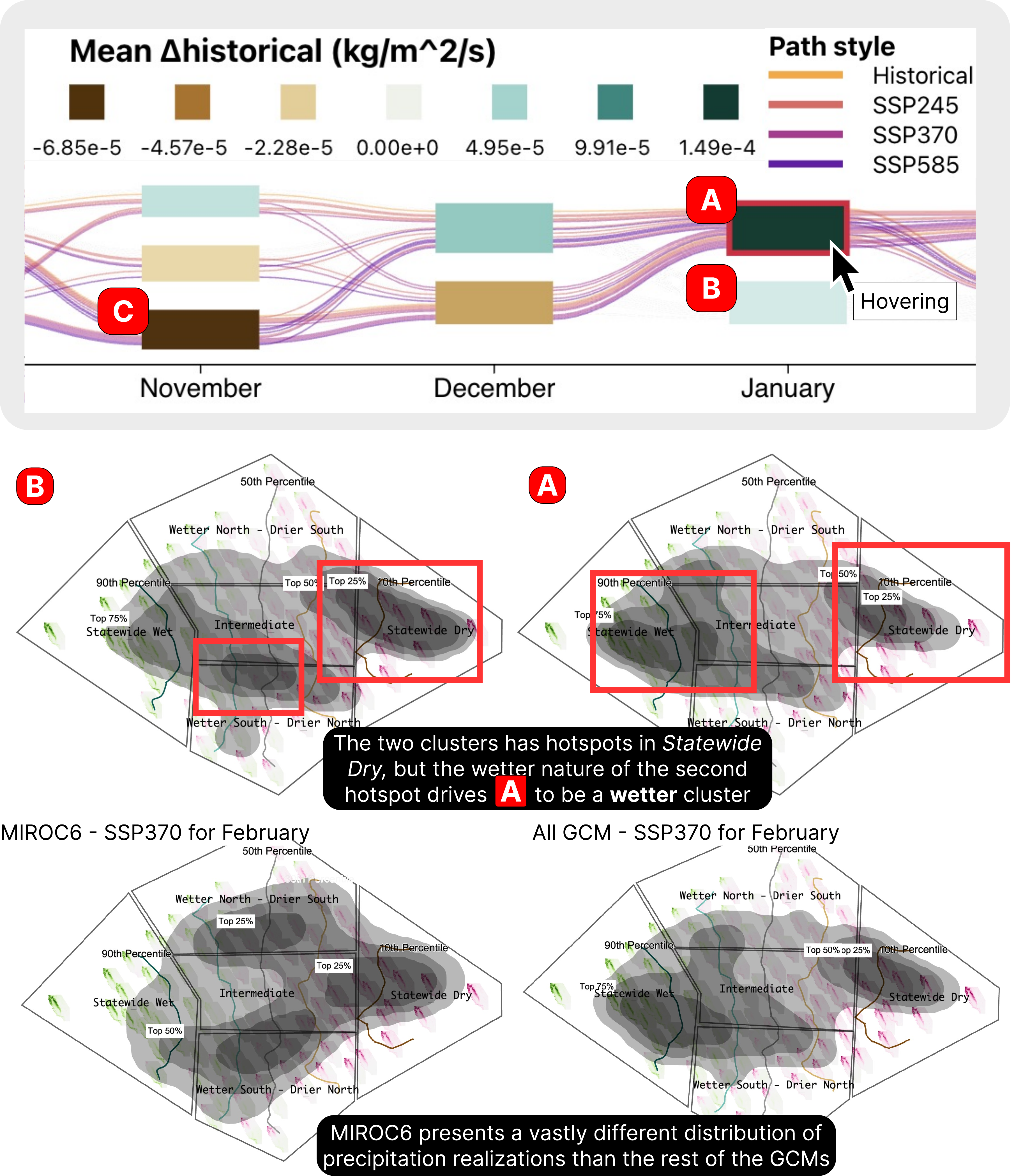}
    \caption{The California case study.% with \ClimateSOM. 
     \textit{The Annotated SOM Node Space} used is shown in~\cref{fig:Step1and2}. The expert explored monthly clustering in model runs in January - here the user has hovered over \dist{A}, displaying the lines that pass through it (\req{R3}). The expert explored the cluster behavior by examining its distributions. Lastly, they also explored a single model run, \textit{MIROC6 - SSP370}, against the same SSP for all GCMs (\req{R2}).}
     \vspace{-0.1in}
    \label{fig:Dan}
\end{figure}

\paragraph{SOM Training}
We experimented with various dimensions of the SOM grid and found that the choice had minimal impact on the results so long they were sufficiently large and selected a 30$\times$30 grid.
We include the impacts of the initial neighborhood radius ratio $kR$ and the neighborhood decay ratio $kS$ on the SOM in the Appendix.
For the California region, we selected $kR=0.03$ and $kS=0.2$ as they provided a good balance explanatory power and SOM grid smoothness.
This provided a SOM grid with 80.5\% of explained variance.
\paragraph{The Case Study}
We outline a case study of the precipitation ensemble over California with the \textbf{E1}, a climate scientist and co-author of this paper.
We will reference~\cref{fig:Dan} for this case study.
First, the expert goes through \step{S1}: Anchor to reorganize the SOM node space to improve the interpretability of the SOM node space.
They place wetter precpitation SOM nodes to the left, dryer nodes to the right, and allow the vertical axis to represent the North-South gradient of the region, placing nodes with higher precipitation in the North on the upper side of the SOM Node Space and vice versa.
\revadd{The LLM-integration reported the area within the \textit{Adjusted SOM Node Space} with negative and positive precipitation over the counties over Southern and Northern California, indicating a presence of SOM nodes with positive precipitation in Southern California and negative precpitation overall.}
Next, \revdel{as they are familiar with the SOM nodes and their precipitation patterns}\revadd{along with manual inspection of the space}, they perform \step{S2}: Annotate producing five annotations regions: (1) \textit{Statewide Wet}, (2) \textit{Wetter North - Drier South}, (3) \textit{Wetter South - Drier North}, (4) \textit{Intermediate}, and (5) \textit{Statewide Dry} (\cref{fig:workflow}-\subimage{3}).
The first question the expert wishes to answer is on the behavior of model runs that predict a wetter California in January and whether there are trends that these model run exhibit in the previous months.
To this end, the expert examines the model run clustering in the circuit-line diagram and identifies two clusters of model runs in January (\req{R3}), one representing a wetter California than the other based on the color of the cluster boxes.

They then compare the two January clusters (\req{R2}), learning that the difference in the two clusters [\dist{A},\dist{B}] is primarily in the density of \textit{Statewide Wet} realizations.
Interestingly, the \textit{Statewide Dry} realizations are both prominently present in both clusters.
They visually confirm that the additional occurrences of \textit{Statewide Wet} realizations in one cluster is the driving difference between the two clusters.
%
% The expert also studies the behavior of the two clusters in December, noticing the difference here is in density in the \textit{Statewide Dry} (\req{R2}).
%
Further, they learn from the circuit-line diagram that the both the drier and wetter cluster in December evolve to a wetter California in January, but realizes that it primarily evolves from a dry cluster in November considering the number of lines that pass through the respective clusters.
Following the dry cluster in November \revadd{[\dist{C}]}, they learn that this cluster is represented by heavy \textit{Statewide Dry} realizations as well as \textit{Wetter South - Drier North} realizations.
Thus the expert concludes that models run that predict a wetter California in January are associated with a overall drier November with a particular dryness in the North of the region, but less clear trends in December.

Another exploration that the expert wishes to undertake is of the behavior of MIROC6, a GCM, in the medium-emission scenario, SSP370, relative to the other GCMs under the same emissions scenario.
Exploring through different months, they observe that MIROC6 in February has a distinct pattern compared to the other GCMs in the same emissions scenario.
While the other GCMs over February are distributed across \textit{Statewide Wet} and \textit{Statewide Dry} realizations, MIROC6 introduces heavy \textit{Wetter North - Drier South}  and \textit{Wetter South - Drier North} realizations (\req{R1}).
This difference in mixture of realizations may warrant further investigation into MIROC6 - SSP370 over February to understand the reasons for the model run's distinct behavior.

\subsubsection{LOCA-Downscaled CMIP6 over the Northwestern US}
\begin{figure*}[h]
    \centering
    \includegraphics[width=\textwidth]{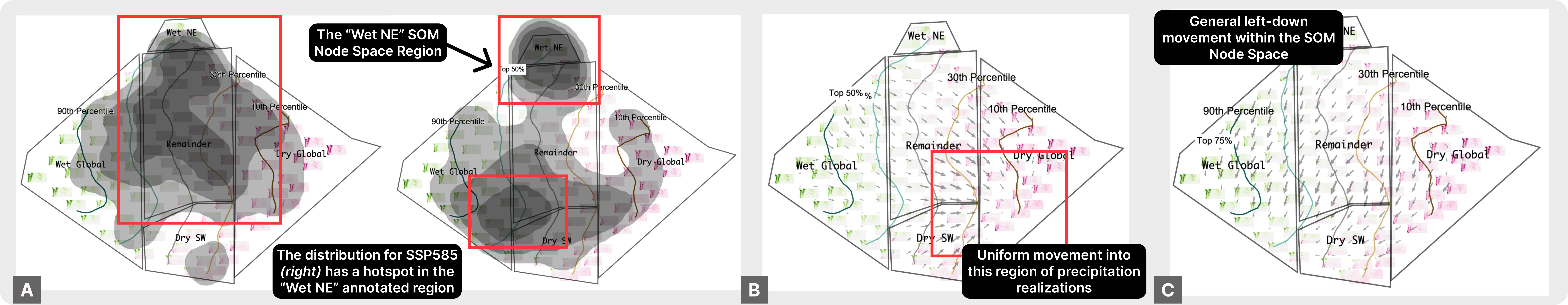}
    \caption{The Northwestern US case study with \ClimateSOM. \highlight{A}: Comparing historical and SSP585 projections in April (\req{R2}), \highlight{B},\highlight{C}: Exploring the two GCM forcing clusters in November and comparing their behavior (\req{R3}).}
    \vspace{-0.3in}
    \label{fig:Tom}
\end{figure*}

% This section recounts a domain expert's experience using \ClimateSOM to analyze LOCA-downscaled CMIP6 precipitation ensemble over the Northwestern US.
We follow the same process to pick the SOM parameters for the Northwestern US case study, selecting $kR=0.03$ and $kS=0.2$ over a $30\times 30$ SOM grid explaining 78.8\% of the variance.
With consultation from an Economist, \textbf{E2}, with experience studying in climate change impacts on the Western Coast of the United States, this case study examines LOCA-downscaled CMIP6 precipitation ensemble over the Northwestern US.
We will refer to ~\cref{fig:Tom} for this case study.
First, the expert goes through \step{S1}: Anchor to reorganize the SOM node space to improve the interpretability of the SOM node space and \step{S2}: Annotate and visually identifies regions of interest, relying on previous domain knowledge of precipitation patterns in the Northwestern US.
This yielded five annotated regions: (1) \textit{Dry Global}, (2) \textit{Wet NE}, (3) \textit{Dry SW}, (4) \textit{Wet Global}, and (5) \textit{Remainder}.
% \begin{figure}[h]
%     \centering
%     \includegraphics[width=0.8\linewidth]{figures/TomSpace.png}
%     \caption{The \textit{The Annotated SOM Node Space} for the Northwestern US case study}
%     \label{fig:TomSpace}
% \end{figure}

% The expert first examines the behavior of all GCMs over the historical period in January noticing the breakdown of the selected months into the annotated regions, \textit{Dry Global} and \textit{Remainder} being the most prominent (\req{R1}).
%
% This is seen in \highlight{B}.
%
% Checking the results for the other Fall months, the expert observes the Fall vs.

First, the expert wishes to compare historical and SSP585 precipitation projections in April. 
Here, this transition is represented by a change from a unimodal distribution of precipitation realizations to a bimodal distribution (\highlight{A}).
In particular, the expert notices a shift to a wetter Northeast (of the region) under this higher emissions future, which they recognize to be a \textit{notable change} with particular significance to agriculture in the region (\req{R2}).
% \begin{figure}[h]
    % \centering
    % \includegraphics[width=\linewidth]{figures/TomBimodal.png}
    % \caption{The shift from a unimodal to bimodal distribution of precipitation realizations in April, comparing historical (left) and SSP585 (right) ensemble model runs.}
    % \label{fig:TomSpace}
% \end{figure}
%
In contrast, in November, the expert observes a movement towards a wetter Northwest (of the region) in a higher emissions future.
The expert therefore confirms that the effect on a higher emissions future on precipitation in the Northwestern US is drastically \textbf{different} between April and November, where April expects higher precipitation in the Northeast compared to the Northwest.
This is an interesting insight, as simple statistical analysis would not reveal such spatial patterns.

% with visual descriptions of the changes.
% \begin{figure}[h]
%     \centering
%     \includegraphics[width=\linewidth]{figures/TomNov.png}
%     \caption{The shift from a precipitation realizations in November, comparing historical (left) and SSP585 (right) ensemble model runs.}
%     \label{fig:TomSpace}
% \end{figure}
%
% Drilling down to a particular climate model, they observe that for ACCESS-CM2 comparing historical to a higher emission model in November, they see an increase in precpitation in the westside and coastal regions of the Northwestern US (\req{R2}).

Next, the expert wishes to explore whether there are outlier GCMs that exhibit different SSP forcing from historical to SSP585 in November. 
Consulting the circuit-line diagram for the GCM clustering of forcings, they observe two clusters of GCMs in November, so they click on the two clusters\revdel{.}\revadd{, inspecting their forcing difference via the vector-field formulation.}
Here, they observe that one of the clusters (\highlight{B}), \textit{``homes in the lower right area of the SOM distribution''} thus representing GCMs that predict a drier November with drier anomalies concentrated toward the Southwest of the region.
On the other hand, the other cluster of GCMs (\highlight{C}), in contrast, collects toward the lower left area of the SOM distribution, and thus a wetter coastal region and wetter border with Canada, highlighting the two clusters of GCMs in its SSP forcing (\req{R3}).
They hypothesize that \revdel{the} the GCM's forcing in the first cluster may be anomalous, and thus may warrant further investigation.
Throughout this session, the expert was impressed with \ClimateSOM's ability to \textit{``visualize big picture patterns quickly''} and that it was \textit{``just not something (they) could do otherwise''}.

\subsection{Evaluating the LLM integration}\label{ssec:llm_eval}
While past works demonstrate the ability of LLMs to generate SQL queries and a general text summarizer, past works on LLMs have not explicitly evaluated two spatial abilities: (1) the spatial ability to map region identifiers to a list of counties and (2) the spatial ability to summarize a list of counties to a short summary.
Additionally, given the possible ambiguity that may arise in how places are defined (i.e., \textit{Southern California} may mean different boundaries to different people), we conducted a short evaluation of our LLM in the above two tasks with 8 participants all familiar with the geography of California.
%
% Specifically, we designed two tasks. 

First, given a common region identifier (e.g., \textit{Southern California}), we asked the participants to evaluate how reasonable the list of counties generated by the LLM was.
Second, given a list of counties randomly selected from one of the textual summaries of a SOM node in the California case study, we asked the participants to evaluate how reasonable the regional summaries were.
Both tasks were evaluated on a 5-point Likert scale. \textit{(1: Strongly Agree, 2: Agree, 3: Neither Agree nor Disagree, 4: Disagree, 5: Strongly Disagree}).
We show examples of the tasks in the Appendix.

Collecting 10 responses of each tasks (80 responses per task), we found that for T1, the average score was 2.11 (SD=1.32) and for T2, the average score was 2.5 (SD=1.55).
This suggests that while our LLM was not perfect, they are able to generate reasonable lists of counties and summaries of regions that most participants found reasonable.
Admittedly, in the first task there were regions where responses were more polarized, especially for vaguely defined region identifiers like \textit{Desert Region of California}; and in the second task, when the list of counties were highly disconnected geographically, generating textual summaries of such regions was more challenging.
Another limitation of the current implementation is that we use counties as the unit of analysis. 
However, natural language descriptions of regions do not always conform to county boundaries--some regional identifiers may span or divide multiple counties. 
As a result, generating \revdel{finer-grained}\revadd{finer} boundaries and summaries that account for these irregularities was difficult.

Lastly, we also test the LLM's PostGIS generation for \ClimateSOM by test generating 20 queries for each of the three categories and manually inspecting the results.
We found that our LLM's generation was accurate in \textasciitilde 90\% of the cases, with the remaining \textasciitilde 10\% being cases where the LLM generated a query with incorrect syntax.
\textbf{In all, we found the failure cases with our LLM integration arising mainly from ambiguous region identifiers for spatial tasks, the county-level unit of analysis, and incorrect syntax generation for the PostGIS queries}.
A conversational interface that can allow the user to correct the LLM when they generate unreasonable results is a future direction.

%% file: tex/7_feedback.tex
In addition to the two experts, \textbf{E1} and \textbf{E2}, who provided the two case studies, we also solicit feedback on \ClimateSOM with a semi-structured interview from four other domain experts, \textbf{E3} - \textbf{E6}, (total of six) by presenting the case studies and the system to them.
These experts range in years of experience from 12 to 48 years in climate science and all have worked with climate ensemble datasets.
We summarize and discuss the feedback on \ClimateSOM below.

\paragraph{On the LLM integration} Since the case studies were conducted by experts who had already become familiar with the SOM nodes, thus able to rely on their past knowledge to provide annotations, we instead asked experts to test the potential of LLM integration.
The experts found the LLM integration particularly useful in certain circumstances. 
\textbf{E2} and \textbf{E3} agreed that LLM integration would be useful if the user had a specific administrative region of their interest.
\textbf{E1} suggested that the integration would be useful if the geospatial region in question was unfamiliar to them.
When visual inspection of the SOM nodes alone is insufficient to understand particular patterns in the SOM Node Space, experts \textit{agreed} that the LLM can be used to both identify regions of interest and to provide textual summaries of the nodes in those regions.
However, \textbf{E3} questioned the reliability of LLMs in providing sensemaking of the SOM Node Space, suggesting visual inspection of the SOM Node Space may be more reliable.
A conversational interface for the LLMs to provide a way to error-correct is a future direction.

\paragraph{On the distributional nature of our data abstraction}
Throughout the case studies, we saw examples of how the distributional nature of our data abstraction of model runs in \ClimateSOM revealed important patterns.
Crucially, as \textbf{E1} agreed, providing a distributional view of model runs above an interpretable space via the steerable DR, annotations, and LLM-empowered sensemaking, experts can quickly identify and understand crucial characteristics of model runs.
%
% Combined with the interpretable SOM Node Space that \ClimateSOM provides via the steerable DR, annotations, and LLM-empowered sensemaking, experts can quickly identify and understand crucial characteristics of model runs. 
%
\textbf{E6} showed significant interest in the descriptive power of our representation, in particular of the vector field formulation of comparison, and noted that they had \textit{``never seen anything like it before.''}
Experts agreed that this distributional nature of data abstraction is a \textbf{novel way to visualize climate ensemble datasets} and a distinguishing feature of \ClimateSOM.

\paragraph{\ClimateSOM can quickly discover novel and meaningful insights} Most importantly, \textit{all experts} agreed that the insights our system uncovers in climate ensemble datasets are valuable—either because they can be identified or because identifying them manually would require a very substantial effort.
%
% Most importantly, \textit{all experts} agreed that the insights that our system can uncover into the climate ensemble datasets are valuable, either in that they can be identified or that it would take substantial manual effort to identify them otherwise.
%
\textbf{E2} remarked that, to generate similar insights \textit{``(their) other methods would involve generating a very large number of static plots and sifting through them.''}
% . and that they therefore \textit{`like this kind on interactive interface.`''}
%
\textbf{E1} and \textbf{E6} also noted the ability to investigate the monthly evolution of the clustering results in \ClimateSOM as a novel and valuable feature.
\textbf{E4} was impressed \revdel{by \ClimateSOM} and suggested that it could also be used to answer other important questions for climate models, such as whether downscaled coarse-resolution global climate models can capture the characteristics of the climate system by comparing it with high-resolution regional climate models.
The feedback gives us confidence that our workflow can generate novel insights into climate ensemble datasets of climatological significance.

\section{Limitations and Future Work}
Despite the overall very positive feedback, experts also pointed out limitations.
First, some experts noted the \textbf{visual complexity of the workflow and interface}, though they positively reviewed the memorability of the interface (i.e., there is a learning curve to the system, but once learned, it is easy to remember how to use it).
Granted, the scope of tasks that \ClimateSOM attempts to address is large and complex, and thus future work will focus on improving the usability of the system with better onboarding tutorials.
Second, compared to common statistical or analytic methods, \ClimateSOM provides \textbf{no native metrics of statistical significance} in its findings.
Statistical tests are commonly used in climate science to assert anomalous patterns, for example, and the inclusion of such metrics within the \ClimateSOM framework is a future direction.
Third, although scalability is not a concern for the Model Run Inspection Views, \textbf{the scalability of the Clustering View} can become a concern when considering datasets with more model runs.
An alternative encoding design to accommodate larger number of model runs is future work. 
\revdel{Lastly}\revadd{In addition}, as introduced in~\cref{sec:problem}, the abstraction in \ClimateSOM \textbf{ignores the temporal ordering of the model runs} which is of interest to many climate scientists.
\revadd{Finally, in both case studies, SOM training took \textasciitilde10 minutes. However,  \textbf{training time will grow linearly with the grid size} potentially making it difficult to find appropriate parameters $kR$ and $kS$. Furthermore, especially at large grid sizes, Euclidean distance may not be an appropriate metric\cite{kappe2022visual} to compare climate scalar fields, and thus we face this limitation.}

%% file: tex/9_conclusion.tex
In this paper, we introduce \ClimateSOM, a novel visual analysis workflow for climate ensemble datasets.
Past works show promise for visual analytics to unravel such complex datasets, but they do not simultaneously address the three key tasks, exploration of a single set of model runs, comparison of two sets of model runs, and clustering of model runs and GCMs, that we identify in collaboration with a domain expert.
We leverage self-organizing maps, steerable DR, LLM-empowered query system to address these tasks, and present new insights into a precipitation ensemble dataset over two different \revdel{spatial domains}\revadd{regions}.
%
% In addition, we solicit expert feedback from six scientists with experience studying such datasets and confirmed that \ClimateSOM can indeed discover climatologically significant insights that are novel or would require substantial manual effort to uncover otherwise, so long the trained SOM can \textit{sufficiently} describe the ensemble. 
\revdel{In addition, we}\revadd{We} gathered feedback from six domain experts and confirmed that \ClimateSOM can uncover climatologically significant insights--many of which are novel or would otherwise require substantial manual effort -- provided that the trained SOM \textit{sufficiently} represents the ensemble.
We also discuss the limitations of \ClimateSOM and suggest future directions, in particular, the integration of statistical tests to quantify the significance of the insights generated by \ClimateSOM.
\revdel{We believe that \ClimateSOM is a valuable tool for climate scientists to quickly and effectively explore and analyze climate ensemble datasets.}
We encourage future work in incorporating novel visual analytics techniques to address the unique challenges in climate science.

%% file: tex/appendix.tex
\section{Prompts used in the LLM integration}
\subsection{Forward Direction}
In the forward direction, we presume the user has a SOM node char- acteristic in mind, (e.g.\textit{``Show me SOM nodes with average above 0 over Southern California''}) and we wish to identify a region within the Adjusted SOM Node Space that corresponds to this requested characteristic.
To this end, we want to build an appropriate PostGIS query that matches the user’s request.
Using LangChain, we first detect any spatial region (e.g. \textit{Southern California}) using LangChain Tools.
If so, we first use the \texttt{Region to Counties} prompt to identify the counties in the region.
\tcbset{colback=gray!10, colframe=black, boxrule=0.5pt, arc=2mm}
Once we have the counties, for which strict definitions are available, we obtain the boundaries for the region in the original query.
Next the \texttt{PostGIS Prompt Generation} prompt, along the LangChain Tool Calling that automatically supplies the computed boundarym is used to generate the PostGIS query.
We note that the SQL query generation query prompt is loosely based on LangChain's documentation on SQL generation\cite{langchainSQL}.
Finally, we use the \texttt{Clean Query} prompt to ensure the query is syntactically correct and complete.

\lstset{
    basicstyle=\fontspec{Source Code Pro},
}
\begin{tcolorbox}[title=PROMPT: Region to Counties, width=\linewidth, colframe=navy]
    \begin{lstlisting}[basicstyle=\ttfamily\footnotesize,breaklines=true, columns=fullflexible,showstringspaces=false, keepspaces=true, belowskip=0pt, aboveskip=0pt,breakindent=0pt] 
Given a US region in the West coast, return the names and states of the counties in that region. Region: {{ Region }} \end{lstlisting}
\end{tcolorbox}
\begin{tcolorbox}[title=PROMPT: PostGIS Prompt Generation, width=\linewidth, colframe=navy, breakable]
    \begin{lstlisting}[basicstyle=\ttfamily\footnotesize,breaklines=true, columns=fullflexible,showstringspaces=false, keepspaces=true, belowskip=0pt, aboveskip=0pt,breakindent=0pt] Given an input question, create a syntactically correct one-line PostGIS query to retrieve the required information. Never query for all the columns from a specific table, only ask for the relevant columns in the question. Don't use any prior knowledge about spatial locations. If queried about the whole region take all grid locations. Nodes or members refer to the grid_ids.
Pay attention to use only the column names that you can see in the schema description. Be careful to not query for columns that do not exist. Do no use the lat or lon columns. Always build valid and complete SRID 4326 PostGIS polygons for the region specified in the question.
Example:
Question: grid_ids whose average value in the Southern California higher than 0.5
Answer: 
SELECT grid_id FROM spatial_data WHERE ST_Intersects(geom, ST_GeomFromText(<polygon text>, 4326)) GROUP BY grid_id HAVING AVG(value) > 0.5;
Question: grid_ids with average higher within Central Valley in California than Southern California
Answer:  
 WITH polygon_a_values AS (
SELECT
    grid_id,
    AVG(value) as avg_value_a
FROM your_table
WHERE ST_Contains(ST_GeomFromText(<polygon text>, 4326), geom)

GROUP BY grid_id
),
polygon_b_values AS (
    SELECT
        grid_id,
        AVG(value) as avg_value_b
    FROM your_table
    WHERE ST_Contains(ST_GeomFromText(<polygon text>, 4326), geom)
    GROUP BY grid_id
)
SELECT a.grid_id
FROM polygon_a_values a
JOIN polygon_b_values b
    ON a.grid_id = b.grid_id
WHERE a.avg_value_a > b.avg_value_b;

Only use the following table: {{table_schema}}
Question: {{question}} 
Write an initial draft of the query. Then double check the PostGIS query for common mistake. Produce just the queries. Don't comment on the changes. Changes may include:
    - ST_ functions are used with the right parameters
    - Specify SRID 4326 for all geometries
    - Using NOT IN with NULL values
    - Using UNION when UNION ALL should have been used
    - Using BETWEEN for exclusive ranges
    - Data type mismatch in predicates
    - Properly quoting identifiers
    - Using the correct number of arguments for functions
    - Casting to the correct data type
    - Using the proper columns for joins

Use format:

First draft: <<FIRST_DRAFT_QUERY>>
Final answer: <<FINAL_ANSWER_QUERY>> \end{lstlisting}
\end{tcolorbox}

\begin{tcolorbox}[title=PROMPT: Clean Query, width=\linewidth, colframe=navy]
    \begin{lstlisting}[basicstyle=\ttfamily\footnotesize,breaklines=true, columns=fullflexible,showstringspaces=false, keepspaces=true, belowskip=0pt, aboveskip=0pt,breakindent=0pt] 
Clean the following into an one-line query with no extra text and make sure its a valid PostGIS query and specify SRID 4326: {{to_be_cleaned}}\end{lstlisting}
\end{tcolorbox}

\subsubsection{Example}
\begin{enumerate}
    \item \textit{\textbf{User Query:}} grid\_ids whose average value over Southern California is higher than 0.
    \item Detect \textit{Southern California} as a region that needs to be resolved.
    \item Resolve \textit{Southern California} to the following counties: \begin{lstlisting}[basicstyle=\ttfamily\footnotesize,breaklines=true, columns=fullflexible,showstringspaces=false, keepspaces=true, belowskip=0pt,breakindent=0pt]
Los Angeles County-CA, San Diego County-CA, Orange County-CA, Riverside County-CA, San Bernardino County-CA, Ventura County-CA \end{lstlisting} 
    \item Generate the boundary for \textit{Southern California} by combining the known boundaries for the counties. \begin{lstlisting}[basicstyle=\ttfamily\footnotesize,breaklines=true, columns=fullflexible,showstringspaces=false, keepspaces=true, belowskip=0pt,breakindent=0pt]
MULTIPOLYGON ((-118.881656 34.791231, ...)) \end{lstlisting} 
    \item Construct the PostGIS query.\\ \textit{\textbf{Output:}}\begin{lstlisting}[basicstyle=\ttfamily\footnotesize,breaklines=true, columns=fullflexible,showstringspaces=false, keepspaces=true, belowskip=0pt]
SELECT grid_id FROM spatial_data WHERE ST_Intersects(geom, ST_GeomFromText('MULTIPOLYGON (((-118.881656 34.791231, -117.632011 34.82227, -117.632996 35.797251, -115.648357 35.809211, -114.633487 35.001857, -114.386699 34.457911, -114.131211 34.26273, -114.52868 33.947817, -114.627125 33.433554, -116.085165 33.425932, -116.106168 32.61848, -117.204917 32.528832, -117.571531 33.312302, -118.1259 33.697151, -118.466962 33.725524, -118.557356 33.987673, -119.158963 34.040253, -119.500953 34.326922, -119.442352 34.901274, -118.881656 34.791231)), ((-118.659924 33.50576, -118.359393 33.464922, -118.251379 33.353397, -118.241051 33.313029, -118.250903 33.280465, -118.50368 33.285314, -118.659924 33.50576)), ((-118.654363 33.074952, -118.523727 33.059972, -118.298469 32.794257, -118.538197 32.81277, -118.654363 33.074952)), ((-119.634795 33.285956, -119.519138 33.3337, -119.360702 33.221452, -119.512001 33.167755, -119.634795 33.285956)), ((-119.468064 34.063139, -119.330926 34.065064, -119.296238 33.99724, -119.454479 33.963308, -119.468064 34.063139)))', 4326)) GROUP BY grid_id HAVING AVG(value) > 0; \end{lstlisting}

\end{enumerate}
\subsection{Backward Direction}
In the backward direction, we presume the user has identified a region within the \textit{Adjusted SOM Node Space} and wishes to generate a summary of the region.
To this end, we first evenly sample within the requested \textit{Adjusted SOM Node Space} region.
Given the user-defined cutoffs for \textit{high}, \textit{moderately high}, \textit{neutral}, \textit{moderate}, \textit{moderately low}, and \textit{low}, we extract the set of counties that fall under each category for every sampled SOM node.
Next, for each category for each node, we use the \texttt{Counties to Regions} prompt to generate a concise summary of the counties.
As an example, when supplied with \textit{[`Coos-Oregon', `Curry-Oregon']} it may generate \textit{`Southern Coastal Oregon'}.
Finally, we aggregate all this text using the \texttt{Aggregate Summaries} prompt to generate a summary of the region.
\begin{tcolorbox}[title=PROMPT: Counties to Regions, width=\linewidth, colframe=navy]
    \begin{lstlisting}[basicstyle=\ttfamily\footnotesize,breaklines=true, columns=fullflexible,showstringspaces=false, keepspaces=false, belowskip=0pt, aboveskip=0pt,breakindent=0pt] 
You are an expert in geographical analysis and regional summarization. Your task is to group and summarize a list of county names into concise regional descriptions based on their geographical characteristics.
When summarizing, consider:
    - Geographic location (e.g., northern, southern, coastal, inland, mountainous).
    - Common regional identifiers (e.g., `Bay Area,' `Central Valley,' `Southern California').

Format the output as summary of less than 20 words. If counties are widely spread, mention multiple distinct regions. Keep the response clear, concise, and relevant. Do not make any inferences about the regions.

For example, you might respond: `Northern coastal areas and the Central Valley, including some southern inland regions.'

{{ list_of_counties }}\end{lstlisting}
\end{tcolorbox}
\begin{tcolorbox}[title=PROMPT: Aggregate Summaries, width=\linewidth, colframe=navy]
    \begin{lstlisting}[basicstyle=\ttfamily\footnotesize,breaklines=true, columns=fullflexible,showstringspaces=false, keepspaces=true, belowskip=0pt, aboveskip=0pt,breakindent=0pt] 
You are a climate scientist tasked with summarizing climate reports. Scientists have provided observations of regions categorized by varying precipitation levels, formatted as follows:

high_precipitation: [high_precipitation_regions], moderate_high_precipitation: [moderate_high_precipitation_regions], neutral_precipitation: [neutral_precipitation_regions], moderate_low_precipitation:[moderate_low_precipitation_regions], low_precipitation: [low_precipitation_regions].

Your objective is to analyze these observations and identify overall trends in approximately 50 words, focusing on the characteristics of the different precipitation regions. Provide a summary that reflects the proportions of each region category putting focusing on the most consistent trends.  Refrain from making any inferences or assumptions beyond the provided data. 

{{ concatenated_descriptions_dictionary }}\end{lstlisting}
\end{tcolorbox}

\section{Algorithm for computing the vector field formulation of comparison}
\begin{algorithm}
  \caption{Computing the vector field formulation of comparison }\label{alg:bootstrap_comparison}

  \begin{algorithmic}[0]
  %-------------- Input & Output -----------------
  \State \textbf{Input:} Distribution \textbf{$R_1$}, Distribution \textbf{$R_2$} of points in the \textit{Annotated SOM Node Space}, Number of bootstrap samples \textbf{k}, Vector field dimension \textbf{n}
  \State \textbf{Output:} Uniform vector field  \textbf{F} of dimension $\textbf{n}\times \textbf{n}$

  %--------------- for loop -----------------------
  \State \textbf{Initialize:} \textbf{results} = $[]$
  \For{$i \in \{1,2,\ldots, k\}$}
  % \For{\textbf{all} pixel $\textbf{I}(x,y)$}
       % \State Consider a window \textbf{W} of size $s \times s$ around $\textbf{I}(x,y)$
       \State Generate bootstrap samples $R_1',R_2'$ 
       \State $OT= $ 2D optimal transport between $R_1'$ and $R_2'$ 
       \State \Comment{\parbox[t]{.6\linewidth}{$OT[i] = (r_1',r_2')$ s.t. $r_1'\in R_1', r_2'\in R_2'$}}
       \State $targets$ = $\{\}$ \Comment{\parbox[t]{.6\linewidth}{$\forall$ $source \in R_1$ store the list of targets matched $target \in R_2$}}
       % \Statex \hspace{1em} \Comment{$OT[i] = (r_1',r_2')$ s.t. $r_1' \in R_1', r_2' \in R_2'$}
       \For {$(r_1',r_2')\in OT$}
            \State $targets[r_1']$.append($r_2'-r_1'$)
       \EndFor
       \For {source $r_1'\in targets$}
            % \State $\textbf{average\_vector}.append(\text{mean}(\textbf{targets}[r_1']))$
            \State $targets[r_1']$ = $mean$($targets[r_1']$) \Comment{Take the average}
       \EndFor
       % average vector, $target-source$ for each source $r_1'$ in \textbf{targets}
       \State $bootstrap\_sample=$ Interpolate $targets$ onto a regular grid of dimension $n\times n$
       \State \textbf{results}.append($bootstrap\_sample$)
       % \State Compute the anisotropic filter $\textbf{F}_{\textbf{W}(x,y)}$ at the location $(x,y)$

       % \State $\textbf{F}_{\text{\textbf{W}(x,y)}} = \left\{\begin{array}{cl}
       %     \textbf{G}(x,y) & \text{if $\textbf{W}(x,y) \geq \text{Mean}(\textbf{W})$} \\
       %     0               & \text{if $\textbf{W}(x,y) < \text{Mean}(\textbf{W})$}
       %   \end{array}\right.$
       % \State $\textbf{Z}(x,y)$ = $\sum\sum ( \textbf{F}_{\text{\textbf{W}(x,y)}} \circ \textbf{W}(x,y) )$
       % \State Compute the weight $w$
       % \State $w = ( s \times s ) \times \sum\sum \textbf{F}_{\text{\textbf{W}}}$
       % \State $w = \tfrac{1}{w}$
  \EndFor

  %----------- Remaining text ----------------
  \State Return the average of \textbf{results} as \textbf{F} 

  \end{algorithmic}
\end{algorithm}
\section{Example images used in the LLM integration evaluation}
\begin{figure}[H]
    \centering
    \includegraphics[width=\linewidth]{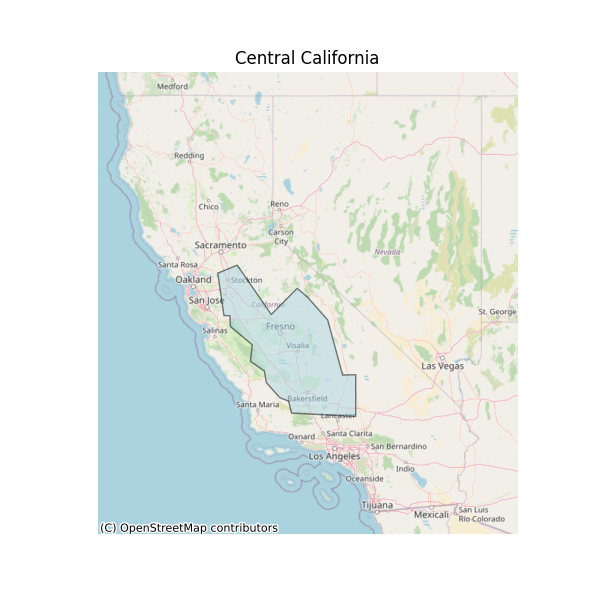}
    \caption{Users were asked to rate in a 5-point likert scale how reasonable the LLM-generated boundary of \textit{Central California} was. Question: \textit{``How reasonable is the above LLM-generated boundary for `Central California'?''}}
\end{figure}

\begin{figure}[H]
    \centering
    \includegraphics[width=\linewidth]{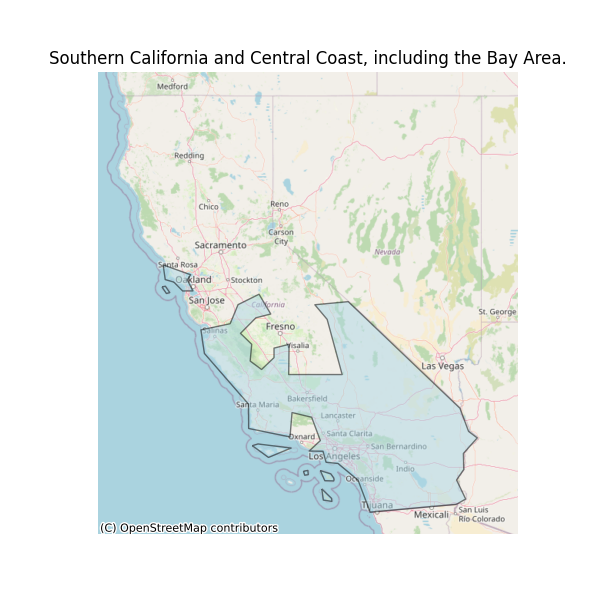}
    \caption{Users were asked to rate in a 5-point likert scale how reasonable the LLM-generated summary of drawn boundary was. Question: \textit{``How reasonable is the above LLM-generated summary in the map title for the drawn boundary?''}}
\end{figure}

% \section{Clustering View for the GCM Clustering by its forcings}

% \begin{figure}[H]
%     \centering
%     \includegraphics[width=\linewidth]{figures/GCMClustering.pdf}
%     \caption{The Clustering View for the GCM clustering variant. The visual encoding is identical to the model run Clustering View, using MDS at each month to place the GCM clusters vertically for every month.}
% \end{figure}

\section{Details of the SOM training for \ClimateSOM}

\subsection{On the metrics for SOM training}\label{ssec:som_training_metrics}
Past research have employed various metrics to evaluate the quality of a SOM \cite{forest2020survey}.
Generally, metrics for SOM are divided into those that evaluate error that the SOM approximation introduces and those that evaluate the topological properties of the SOM grid.
\ClimateSOM requires both classes of metrics to measure that the trained SOM abstraction retains \textit{sufficient} information whilst also ensuring that the MDE dimensionality-reduced space is connected and interpretable.
The most common metrics for the respective categories are quantization error ($QE$) and topographic error ($TE$).
%
% For a given dataset $D=\{d_1, d_2, \ldots, d_n\}$, the quantization error is defined as the average distance between each input field $d_i$ and its BMU, $BMU(d_i)$.
%
% Topographic error is defined as the proportion of all data for which the first and second BMUs are \textit{not} adjacent in the SOM grid.
%
% While these are useful metrics, we introduce two more metrics that we have found useful for \ClimateSOM: (1) Explained Variance ($EV$) and (2) Variation of Smoothness ($VS$).
While these are useful metrics, we present two more metrics that we have found useful for \ClimateSOM: (1) Explained Variance ($EV$) and (2) Mean Smoothness ($\mu_S$).

\noindent
\textbf{Explained Variance}, akin to its definition in principal components analysis (PCA), measures the proportion of variance in the input data set that is explained by the SOM grid, where 0 indicates no explanatory power and 1 indicates full explanatory power.
%%
% Formally, for a data set $D$ the explained variance is defined for a trained SOM is defined:
%\begin{equation}
%    EV = 1 - \frac{\sum_{i=1}^{n}\sum_{j=1}^{m}(d_{i_j} - (BMU(d_i))_j)^{2}}{\sum_{i=1}^{n}\sum_{j=1}^{m}(d_{i_j} - (\bar{d})_j)^{2}}
%\end{equation}
%%
%where $n$ is the number of members in $D$, $m$ is the number of spatial locations, $d_{i_j}$ is the $j^{th}$ element of the $i^{th}$ input field, $BMU(d_i)_j$ is the $j^{th}$ element of the BMU of the $i^{th}$ input field, and $\bar{d}_j$ is the $j^{th}$ element of the mean of the input data set.
%% 
We replace $QE$ with Explained Variance to provide a unitless measure of the explanatory power of the SOM grid.

%\textbf{Variation of Smoothness} is another metric that we introduce to more carefully evaluate the smoothness of the SOM grid.
%%
%Given a SOM grid $G$, we define local smoothness, $LS$, at any grid node $g_{i,j}$ as the average Euclidean distance between $g_{i,j}$ and its four immediate neighbors.
%%
%Finally, considering the standard deviation of all $g_{i,j}$, we define the Variation of Smoothness as $\sigma_{LS}/ \mu_{LS}$.
%%
%Lower values of Variation of Smoothness indicate a more consistent level of smoothness across the grid and vice versa.
%%
%We found that topographic error was not sufficient to capture the required smoothness of the SOM grid to ensure the interpretablitity of the MDE result and thus replace the topology-based error metric with the Variation of Smoothness.

%In all, SOM training for \ClimateSOM is the balancing of its explanatory power ($QE,EV$) and smoothness ($VS$).
%%
%The following, \cref{ssec:som_training_parameters}, discusses the choice of parameters in the context of these metrics and the implications.
\noindent
\textbf{Mean Smoothness} is another metric that we introduce to more carefully evaluate the smoothness of the SOM grid as non smoothness can produce disconnected MDE results.
Given a SOM grid $G$, we define local smoothness, $LS$, at any grid node $g_{i,j}$ as the average Euclidean distance between $g_{i,j}$ and its four immediate neighbors and define the Mean Smoothness as $\mu_{g_{i,j}}$.
We found that topographic error was not sufficient to capture the required smoothness of the SOM grid to ensure the interpretablitity of the MDE result and thus replace the topology-based error metric with Mean Smoothness.

In all, SOM training for \ClimateSOM is the balancing of its explanatory power ($QE,EV$) and smoothness ($\mu_S$).
The following, \cref{ssec:som_training_parameters}, discusses the choice of parameters in the context of these metrics and the implications.

\subsection{On the parameter choices for SOM training}\label{ssec:som_training_parameters}
\begin{figure}[h]
    \centering
    \includegraphics[width=\linewidth]{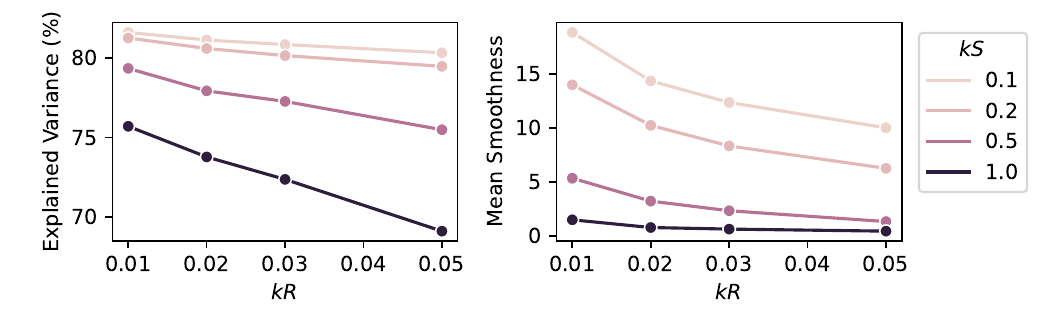}
    \caption{Effects of SOM Parameters, $kR$ and $kS$ on the explanatory power and smoothness of the SOM grid trained for the California Use Case}
    \label{fig:SOMParameters}
\end{figure}

Among the parameters required for SOM training the initial neighborhood size $(kR)$ and the neighborhood size decay $(kS)$ are most important and we thus list their influences here.
\begin{itemize}[left=5pt, itemsep=0pt, labelsep=1em, align=left]
    % \item[\textbf{dim}] The dimension of the SOM grid defines the number of nodes \ClimateSOM uses to approximate the variations in the input ensemble. Larger values tend to result in better explanatory power, but adds complexity to the MDE. While non-square grids can be used, we opt for square grids for simplicity.
    \item[\textbf{kR}] The Gaussian neighborhood function depends on the parameter $\sigma$. The term $kR$ defines its initial value as the ratio between the SOM grid area and the neighborhood update area, given by $kR=(\sigma_{initial})^2/ dim^2$. 
        %
        % Generally, a larger $kR$ results in a smoother SOM grid with less explanatory power (i.e. low $VS$, low $EV$ and high $QE$) and a smaller $kR$ results in the opposite (i.e. high $VS$, high $EV$ and low $QE$).
        Generally, a larger $kR$ results in a smoother SOM grid with less explanatory power and a smaller $kR$ results in the opposite.
    \item[\textbf{kS}] In order for the SOM to \textit{fit} the data, it is common to linearly decay the neighborhood size over time \cite{gibson_use_2017}. %
        $kS$ defines the final value of the neighborhood size as a ratio of the initial value, $kS=\sigma_{final}/ \sigma_{initial}$.
        Similar to $kR$, $kS$ also expresses a trade-off between smoothness and explanatory power.
        A larger $kS$ results in a smoother SOM grid with less explanatory power and a smaller $kS$ results in the opposite.
\end{itemize}
Other parameters such as the dimensions of the SOM grid, as long as they were sufficiently large to enable sufficient explained variance, learning rate decay, the number of iterations, and the initialization method are part of the training process but were found less critical to the performance, and hence we omit them from discussion.
\cref{fig:SOMParameters} shows the tradeoff between $kR$ and $kS$ as it relates to the explained variance and mean smoothness -- choosing the right parameter is a balance act between the two.

\section{Details of the ensemble members used in this work}
\begin{table*}
    \centering
    \begin{tabular}{|c|c|c|}
        \hline
        Global Climate Model (GCM) & Shared Socioeconomic Pathway (SSP) & Variant \\
        \hline
        \hline
        ACCESS-CM2 & historical & r1i1p1f1 \\
        CESM2-LENS & historical & r1i1p1f1 \\
        CNRM-ESM2-1 & historical & r1i1p1f2 \\
        EC-Earth3-Veg & historical & r1i1p1f1 \\
        EC-Earth3 & historical & r1i1p1f1 \\
        FGOALS-g3 & historical & r1i1p1f1 \\
        GFDL-ESM4 & historical & r1i1p1f1 \\
        HadGEM3-GC31-LL & historical & r1i1p1f3 \\
        INM-CM5-0 & historical & r1i1p1f1 \\
        IPSL-CM6A-LR & historical & r1i1p1f1 \\
        KACE-1-0-G & historical & r1i1p1f1 \\
        MIROC6 & historical & r1i1p1f1 \\
        MPI-ESM1-2-HR & historical & r1i1p1f1 \\
        MRI-ESM2-0 & historical & r1i1p1f1 \\
        TaiESM1 & historical & r1i1p1f1 \\
        \hline

        ACCESS-CM2 & ssp245 & r1i1p1f1 \\
        CESM2-LENS & ssp245 & r1i1p1f1 \\
        EC-Earth3-Veg & ssp245 & r1i1p1f1 \\
        EC-Earth3 & ssp245 & r1i1p1f1 \\
        FGOALS-g3 & ssp245 & r1i1p1f1 \\
        GFDL-ESM4 & ssp245 & r1i1p1f1 \\
        HadGEM3-GC31-LL & ssp245 & r1i1p1f3 \\
        INM-CM5-0 & ssp245 & r1i1p1f1 \\
        IPSL-CM6A-LR & ssp245 & r1i1p1f1 \\
        KACE-1-0-G & ssp245 & r1i1p1f1 \\
        MIROC6 & ssp245 & r1i1p1f1 \\
        MPI-ESM1-2-HR & ssp245 & r1i1p1f1 \\
        MRI-ESM2-0 & ssp245 & r1i1p1f1 \\
        TaiESM1 & ssp245 & r1i1p1f1 \\

        \hline
        ACCESS-CM2 & ssp370 & r1i1p1f1 \\
        CESM2-LENS & ssp370 & r1i1p1f1 \\
        CNRM-ESM2-1 & ssp370 & r1i1p1f2 \\
        EC-Earth3-Veg & ssp370 & r1i1p1f1 \\
        EC-Earth3 & ssp370 & r1i1p1f1 \\
        FGOALS-g3 & ssp370 & r1i1p1f1 \\
        GFDL-ESM4 & ssp370 & r1i1p1f1 \\
        INM-CM5-0 & ssp370 & r1i1p1f1 \\
        IPSL-CM6A-LR & ssp370 & r1i1p1f1 \\
        KACE-1-0-G & ssp370 & r1i1p1f1 \\
        MIROC6 & ssp370 & r1i1p1f1 \\
        MPI-ESM1-2-HR & ssp370 & r1i1p1f1 \\
        MRI-ESM2-0 & ssp370 & r1i1p1f1 \\
        TaiESM1 & ssp370 & r1i1p1f1 \\
        \hline

        ACCESS-CM2 & ssp585 & r1i1p1f1 \\
        CESM2-LENS & ssp585 & r1i1p1f1 \\
        EC-Earth3-Veg & ssp585 & r1i1p1f1 \\
        EC-Earth3 & ssp585 & r1i1p1f1 \\
        FGOALS-g3 & ssp585 & r1i1p1f1 \\
        GFDL-ESM4 & ssp585 & r1i1p1f1 \\
        HadGEM3-GC31-LL & ssp585 & r1i1p1f3 \\
        INM-CM5-0 & ssp585 & r1i1p1f1 \\
        IPSL-CM6A-LR & ssp585 & r1i1p1f1 \\
        KACE-1-0-G & ssp585 & r1i1p1f1 \\
        MIROC6 & ssp585 & r1i1p1f1 \\
        MPI-ESM1-2-HR & ssp585 & r1i1p1f1 \\
        MRI-ESM2-0 & ssp585 & r1i1p1f1 \\

        \hline
    \end{tabular}
    \caption{LOCA-Downscaled CMIP6 Precipitation Ensemble members used in this work}
\end{table*}